\documentstyle[epsfig,epsf,eqsecnum,aps,12pt]{revtex}
\def\beq{\begin{equation}}
\def\eeq{\end{equation}}

\def\bea{\arraycolsep .1em \begin{eqnarray}}
\def\eea{\end{eqnarray}}

\def\lsim{\mathrel{\lower4pt\hbox{$\sim$}}\hskip-12.5pt\raise1.6pt\hbox{$<$}\;}
\def\gsim{\mathrel{\lower4pt\hbox{$\sim$}}
\hskip-12.5pt\raise1.6pt\hbox{$>$}\;}

\def\eq#1{(\ref{#1})}

\def\s0#1#2{\mbox{\small{$ \frac{#1}{#2} $}}}
\def\0#1#2{\frac{#1}{#2}}

\begin{document}

\begin{center}

\thispagestyle{empty}

{\normalsize\begin{flushright}CERN-TH-2001-132\\[12ex] 
\end{flushright}}

\mbox{\large \bf 
Photon Self-Energy in a Color Superconductor}\\[6ex]

{
Daniel F.~Litim\footnote{E-Mail:  Daniel.Litim@cern.ch} and
Cristina Manuel\footnote{E-Mail:  Cristina.Manuel@cern.ch}}
\\[4ex]
{\it  Theory Division, CERN, CH-1211 Geneva 23, Switzerland.}
\\[10ex]
 
{\small \bf Abstract}\\[2ex]
\begin{minipage}[t]{14cm} 
 {\small 
  
   In   a   color  superconductor   the   diquark  condensates   break
   spontaneously both the color and ordinary electromagnetism, leaving
   a   remaining  rotated   $U(1)$  symmetry   unbroken.    The  gauge
   interactions associated to this  rotated symmetry may be considered
   as the in-medium electromagnetism.  We compute the in-medium photon
   self-energy in the presence of diquark condensates at high baryonic
   density and weak  coupling. This is done to  one-loop order for the
   cases of two and three  quark flavors. For vanishing temperature, a
   detailed discussion  is given of  the low momentum behavior  of the
   photon  polarization  tensor. A  simple  physical  picture for  the
   propagation of  light in color  superconducting media is  obtained. 
   The main new  effect is due to the  diquark condensates, which lead
   to  a  strong  dielectric  constant  of the  medium.  The  magnetic
   permeability  remains unchanged,  because  the primary  condensates
   have vanishing spin  and angular momentum. In the  two flavor case,
   an  additional  contribution  arises  due  to  gapless  quarks  and
   electrons,  which is  responsible for  Debye screening  effects. We
   also discuss the  low energy effective theory for  the three flavor
   case in the presence of electromagnetic interactions.

}

\end{minipage}
\end{center}

\newpage
\pagestyle{plain}
\setcounter{page}{1}

\section{Introduction}
\label{Intro}

Quantum Chromodynamics  (QCD) in the  regime of high  baryonic density
and low temperature displays the phenomenon of color superconductivity
\cite{Barrois:1977xd,Bailin:1984,Alford:1998,Alford:1999mk}        (see
\cite{Rajagopal:2000wf} for  a recent review).  This  is a consequence
of Cooper's theorem, which states that an attractive interaction close
to the Fermi  surface makes the system unstable  towards the formation
of  condensates.  In  QCD the  attractive interaction  is  provided by
one-gluon exchange  between quarks in  a color antisymmetric  channel. 
The existence  of the  condensates leads to  energy gaps for  both the
fermionic quasiparticles and some or  all gluons.  It is expected that
a color  superconducting state  of matter could  exist in the  core of
compact stars.  Some basic properties  of very dense  stellar objects,
like neutron  stars, may be understood  through the study  of cold and
dense quark matter.

At  asymptotically high  baryonic density,  and because  of asymptotic
freedom of QCD, it is possible  to compute the quark gap and the gluon
masses               from               first               principles
\cite{Son:1999,Schafer:1999b,Pisarski:2000bf,Hong:2000tn,Brown:2000aq,Manuel:2000nh}. 
At present, it is unclear to what extent these results can be extended
to   more   moderate   values   of   the   baryonic   density,   where
non-perturbative effects may  become relevant \cite{Rajagopal:2000rs}. 
Furthermore,  there  are  no  reliable methods  available  to  perform
numerical simulations of QCD  at finite chemical potential. Therefore,
results obtained at asymptotically  large baryonic densities cannot be
compared with those for less dense  systems. In this light, it is most
important  to  provide a  sound  understanding  of  the weak  coupling
regime.  We expect that these  studies lead to both the qualitatively,
and probably  also semi-quantitatively, correct behavior  of the color
superconductor for densities attained in the core of neutron stars.
 
The diquark condensates of a color superconductor break spontaneously,
partially or totally, the non-Abelian gauge symmetry, depending on the
number of  quark flavors participating  in the condensation.   For two
light quark flavors  (2SC phase), the color group  $SU(3)_c$ is broken
down to an $SU(2)_c$  subgroup \cite{Alford:1998}. Five gluons acquire
masses through the Anderson-Higgs mechanism, while three gluons remain
massless.  Also,  not all quarks attain  a gap. For  three light quark
flavors  the   condensates  lock   both  color  and   flavor  symmetry
transformations      (color-flavor      locking     or      CFL~phase)
\cite{Alford:1999mk}.  The CFL  condensates  spontaneously break  both
color and  flavor symmetries $SU(3)_c \otimes  SU(3)_L \otimes SU(3)_R
\otimes  U(1)_B  \rightarrow SU(3)_{c+L+R}  \otimes  {\cal Z}_2$.  All
gluons  acquire masses  in this  case, and  all quarks  attain a  gap. 
Furthermore, there is a Goldstone  boson associated to the breaking of
baryon number,  and eight Goldstone bosons associated  to the breaking
of chiral symmetry.

The infrared  physics in  a color superconductor  is dominated  by its
light  degrees of  freedom. For  the two-flavor  color superconductor,
those  are the  unbroken  gauge  fields and  the  gapless quarks.   An
effective theory for the long  distance physics of the $SU(2)_c$ gauge
fields    has   been   recently    been   discussed    for   vanishing
\cite{Rischke:2000cn}        and       non-vanishing       temperature
\cite{Litim:2001je}.  For the CFL color superconductor the light modes
are  the Goldstone  bosons associated  to the  breaking of  the global
symmetries.   At  low  energies  an effective  Lagrangian  and  chiral
perturbation theory can  be used to study the  dynamics of these light
modes
\cite{Casalbuoni:1999wu,Son:2000cm,Rho:2000xf,Hong:2000ei,Manuel:2000wm,Zarembo:2000pj,Beane:2000ms,Miransky:2001bd,Manuel:2001xt}.

The long distance physics in a color superconductor is modified if one
includes                  electromagnetic                 interactions
\cite{Alford:1999mk,Alford:2000pb}.   Both  the  CFL and  2SC  diquark
condensates break spontaneously the standard electromagnetic symmetry.
However,  a linear  combination of  the  original photon  and a  gluon
remains massless in both cases.  This  new field plays the role of the
``in-medium'' photon in the superconductor.

In the present article, we  compute the photon self-energy for the CFL
and 2SC phases  at very large densities, in the  weak coupling regime. 
We  show how  the  quark  condensates affect  the  propagation of  the
``rotated'' photon.  In the  literature, it has sometimes been assumed
that the photon  propagates as in the vacuum, or as  in a dense medium
without  Cooper pairs  of  quarks.  We  will  see that  none of  these
assumptions holds  true.  Our results  are, therefore, relevant  for a
number  of properties of  color superconducting  quark matter.  Let us
mention two  examples. The  first one concerns  the charged  pions and
kaons of the CFL phase  which acquire a mass of electromagnetic origin
even  in  the  chiral  limit   $m_q  =0$,  as  first  pointed  out  in
\cite{Hong:2000ng}. A computation  of this electromagnetic mass, which
requires the  evaluation of a  three-loop diagram with  dressed photon
propagators, is still missing, although  a rough estimate was given in
\cite{Manuel:2001xt}. A  second example concerns  transport properties
of superconducting quark matter.  Knowing the infrared behavior of the
electromagnetic interactions  is essential for  a reliable computation
of transport coefficients such  as viscosities, electrical and thermal
conductivities of the superconducting medium.

The paper  is organized as follows. In  Sect.~\ref{3-flavor}, we study
the photon  self-energy for the three flavor  color superconductor. We
review  the   symmetry  breaking   pattern  induced  by   the  diquark
condensates  in  a   three-flavor  color  superconductor.   Using  the
non-linear   framework,   the  unbroken   $U(1)$   symmetry  and   the
corresponding in-medium photon are  identified.  We then study how the
rotated  photon couples  to all  the charged  particles of  the system
(Sect.~\ref{3-electro}). The  Feynman rules for  quark propagators and
vertices are given (Sect.~\ref{3-Feynman}), and the quark contribution
to the  one-loop photon self-energy  is computed (Sect.~\ref{3-traces}
and  Sect.~\ref{3-sum}).  We  discuss  the photon  self-energy in  the
infrared region  and emphasize the most relevant  modifications due to
the diquark  condensates on the  propagation properties of  the photon
(Sect.~\ref{3-self}).   The low  energy  effective theory  of the  CFL
phase,   including   electromagnetic   interactions,  is   constructed
(Sect.~\ref{3-low}).  We close with  a discussion of the main physical
picture     of     photon     propagation     in     the     CFL~phase
(Sect.~\ref{3-Discussion}).   In  Sect.~\ref{2-flavor}  we repeat  the
same analysis for the  two-flavor color superconductor.  We begin with
the  discussion of  the  in-medium  photon and  its  couplings to  all
charged    particles   in    a    two-flavor   color    superconductor
(Sect.~\ref{2-electro}). The  Feynman rules for  quark propagators and
vertices are given (Sect.~\ref{2-Feynman}), and the computation of the
quark  contribution to  the one-loop  photon self-energy  is performed
(Sect.~\ref{2-traces}). The  low-momentum limit of  the self-energy is
studied (Sect.~\ref{2-self}), and the  main physical picture of photon
propagation in the  2SC~phase is discussed (Sect.~\ref{2-Discussion}). 
Sect.~\ref{Summary}  closes  with a  summary  of  the  results and  an
outlook.   In  the  Appendix,  we  compute  the  contribution  to  the
self-energy due  to charged pions  and kaons in  the CFL phase.  It is
shown that charged mesons are negligible for the photon self-energy in
the infrared limit. Throughout we  work with natural units, $\hbar = c
= k_B =1$.

\section{Photon self-energy in the three-flavor color superconductor}
\label{3-flavor}

\subsection{Electromagnetic interactions}\label{3-electro}

The  ground state of  QCD at  high baryonic  density with  three light
quark   flavors  is   described   by  the   (spin  zero)   condensates
\cite{Alford:1999mk}
\beq
\label{eq:2.1}
  \langle q^{ai}_{L } q^{bj}_{L }\, \rangle 
=-\langle q^{ai}_{R } q^{bj}_{R }\, \rangle 
=   k_1 \delta^a_i \delta^b_j
  + k_2 \delta^a_j \delta^b_i \ ,
\eeq
where  $q_{L/R}$  are Weyl  spinors  (a  sum  over spinor  indices  is
understood),  and $a,b$  and $i,j$  denote flavor  and  color indices,
respectively.  These CFL condensates break spontaneously color, chiral
and  baryon number  symmetries. As  a  result, all  the gluons  become
massive  through the  Anderson-Higgs mechanism,  while there  are nine
Goldstone bosons associated to the breaking of the global symmetries.

The diquark condensates also break spontaneously the standard
electromagnetic symmetry.  However, a combination of the
electromagnetic generator and a $SU(3)$ generator leaves the CFL
ground state invariant \cite{Alford:1999mk}.  Thus, a linear
combination of the original photon and a gluon remains massless and
plays the role of the ``new'' photon in the color superconductor.

In order to explicitly identify the massless linear combination of
gluon and photon it is very convenient to use the non-linear framework
\cite{Casalbuoni:1999wu}.  One first introduces left-handed and
right-handed transforming fields
\begin{equation}
\label{eq:2.2}
L^{ai} \sim \epsilon^{ijk} \epsilon^{abc} 
\langle q^{bj}_L q^{ck }_L \rangle^* \ , 
\qquad  
R^{ai} \sim \epsilon^{ijk} \epsilon^{abc}
 \langle q^{bj}_R q^{ck }_R \rangle^* \ .
\end{equation}
These fields contain the would-be Goldstone bosons which are to be
eaten by the gluons through the Higgs mechanism, as well as the
Goldstone bosons associated to chiral symmetry breaking.  Under
$SU(3)_c \otimes SU(3)_L \otimes SU(3)_R\otimes U(1)_{\rm e.m.}$ they
transform as
\begin{equation}
\label{eq:2.3}
L \rightarrow  U_1\, U_L\,  L\  U_c^{\dagger} \ , \qquad 
R \rightarrow  U_1\, U_R\,  R\  U_c^{\dagger} \ .
\end{equation}
Therefore, the covariant derivative acting on these fields is
\begin{equation}
\label{eq:2.4}
D_\mu L =  \partial_\mu L 
         - i e\, Q\,  A_\mu\,  L 
         - i g\,  G_\mu ^n \,  L\,  T^n \ ,
\end{equation}
where $Q = {\rm diag} (2/3, -1/3, -1/3)$ is the quark charge matrix,
$e$ and $g$ are the electromagnetic and strong coupling constants, and
$T^n$ are the $SU(3)$ generators. The covariant derivative acting on
$R$ acts in the same way as for $L$.  The kinetic term
\beq
  \label{eq:2.5}
  {\rm Tr} (D_\mu L^\dagger D^\mu L) 
+ {\rm Tr} (D_\mu R^\dagger D^\mu R)
\eeq
then tells us which are the massive and massless gauge eigenstates in
the theory. The diagonalization of the gauge mass matrix simplifies
upon replacing the standard $SU(3)$ generators by $T^8 =
\frac{\sqrt{3}}{2} Q$ and $T^3 = {\rm diag} (0, 1/2, -1/2)$.

As a result, the gluons $G_\mu ^n$ with $n =1, \ldots, 7$ are all
massive. Their masses, which have been computed from QCD
\cite{Son:2000cm,Rischke:2000ra}, are of the order $\sim g \mu$, with
$\mu$ the chemical potential. In addition, there is a combination of
gluon and photon, ${\widetilde G}^8$, which is massive, while the
orthogonal combination, ${\widetilde A}$, is massless
\cite{Alford:2000pb,Casalbuoni:1999wu,Gorbar:2000ms,Casalbuoni:2000jn}
\footnote{The mixing angle of \cite{Alford:2000pb} differs from ours
  due to a difference in the normalization of the $SU(3)$
  generators.}:
\begin{mathletters}
\label{eq:2.6}
\bea
{\widetilde G}^8_\mu & = & \ \ \, 
 \cos{\theta_{\rm CFL}}\, G^8_\mu + \sin{\theta_{\rm CFL}}\, A_\mu \ ,
\\
{\widetilde A}_\mu & = & 
- \sin{\theta_{\rm CFL}}\, G^8_\mu + \cos{\theta_{\rm CFL}}\, A_\mu \,.
\eea
\end{mathletters}%
Here, the rotation angle $\theta_{\rm CFL}$ is defined as
\beq
\label{eq:2.7}
\cos{\theta_{\rm CFL}} =  \frac{\sqrt{3} g}{\sqrt{3 g^2 + 4 e^2}} \ ,
\qquad  
\sin{\theta_{\rm CFL}} = \frac{2 e}{\sqrt{3 g^2 + 4 e^2}} \ .
\eeq 
The field ${\widetilde A}$ plays the role of the ``rotated'' or
in-medium photon in the superconductor. In contrast to the other
gluons, the mass of ${\widetilde G}^8_\mu$ is of the order $\sim
\sqrt{g^2 +\frac{4}{3} e^2} \,\mu$.

We now turn to the charge eigenstates and their coupling to the
in-medium photon. For the quark matter fields the Lagrangian reads
\bea
\label{eq:2.8}
{\cal L}^{e.m.}_{\rm quarks} =  
{\bar \psi} i {\widetilde D}_\mu \gamma^\mu \psi =  
{\bar \psi} 
\left( i\partial_\mu - {\widetilde e}\, {\widetilde Q}\, 
{\widetilde A}_\mu \right)
\gamma^\mu  \psi \,,
\eea
where we have introduced the gauge coupling of the in-medium photon
$\widetilde e = e \cos{\theta_{\rm CFL}}$. The charge matrix
${\widetilde Q}$ is a matrix in flavor$_{(3\times3)}$ $\otimes$
color$_{(3 \times 3)}$ space
\beq
\label{eq:2.9}
{\widetilde Q} = Q \otimes 1 - 1 \otimes Q \,.
\eeq
The linear combinations 
\beq
\label{eq:2.10}
G_\mu ^{\pm} \equiv \frac{1}{\sqrt{2}} \left(G_\mu^4 \mp i G_\mu^5 
\right) \ , 
\qquad 
H_\mu ^{\pm} \equiv \frac{1}{\sqrt{2}} \left(G_\mu^6 \mp i G_\mu^7
\right) \ , 
\eeq
are $\widetilde Q$-charge eigenstates with charge $\pm \widetilde e$.
They remain mass eigenstates, because the first seven gluons have
equal masses anyway. Taking into account the values of the total
antisymmetric constants of $SU(3)$, $f^{458}=f^{678} = \sqrt{3}/2$, we
find the coupling of the charged gluons to the new photon
\beq
\label{eq:2.11}
{\cal L}^{e.m.}_{\rm gluons} = 
\frac{1}{2}\,
{\widetilde D}^{}_{[\mu } G^{+}_{\nu]}
{\widetilde D}^{}_{[\mu } G^{-}_{\nu]} 
+\frac{1}{2}\,
{\widetilde D}^{}_{[\mu } H^{+}_{\nu]}
{\widetilde D}^{}_{[\mu } H^{-}_{\nu]} \,.
\eeq
Here, ${\widetilde D}_{\mu}X^{\pm}\equiv (\partial_\mu \pm i
{\widetilde e}\, {\widetilde A_\mu}) X^{\pm}$, $X=G$ or $H$, and
${A}_{[\mu} B_{\nu]}\equiv A_\mu B_\nu-A_\nu B_\mu$.  Notice that in
Eq.~(\ref{eq:2.11}) we have written only the part of the Lagrangian
which contains the coupling with the new photon while omitting the
self-interactions amongst gluons, or the mass terms. Those can be
read-off from Eq.~(\ref{eq:2.5}).

Finally, the charged pions and kaons also couple to the new photon
\beq
\label{eq:2.12}
{\cal L}^{e.m.}_{\rm GB} = 
 \widetilde D_{0}  \pi^{+ } {\widetilde D}_{0 } \pi^-  
- v^2_\pi
\widetilde D_{i}  \pi^{+ } {\widetilde D}_{i } \pi^- +  
{\widetilde D}_{0} K^{+ } {\widetilde D}_{0 } K^- 
- v^2_\pi
\widetilde D_{i}  K^{+ } {\widetilde D}_{i } K^-
\ ,
\eeq
where $v_\pi = 1/\sqrt{3}$ is the meson velocity \cite{Son:2000cm}.
Electromagnetic effects also generate a mass term for the charged
pions and kaons, even in the chiral limit $m_q =0$, although its
precise value is yet unknown.\\

\begin{center}
\begin{tabular}{ccccccccccc}
\hline\hline\\[-2ex]
& $G_\mu^1$ & $G_\mu^2$ & $G_\mu^3$ 
& ${}\quad$
& $G_\mu^{+}$      & $G_\mu^{-}$ 
& $H_\mu^{+}$      & $H_\mu^{-}$ 
& ${}\quad$
& $\widetilde G_\mu^8$ 
\\[1ex] \hline\\[-2ex]
$\widetilde Q$-charge ${}\quad$ 
& $\ 0\ $ & $\ 0\ $ &$\ 0\ $ 
&
& $\ 1\ $ & $-1$ 
& $\ 1\ $ & $-1$  
&
& $\ 0\ $ \\[.5ex]
\hline\hline
\end{tabular}
\end{center}
\begin{center}
\begin{minipage}{.6\hsize}
{\small {\bf Table~1}: $\widetilde Q$-charges of gluons in the CFL phase, 
and in units of $\widetilde e = e \cos \theta_{\rm CFL}$.}
\end{minipage}
\end{center}

\begin{center}
\begin{tabular}{ccccccccccc}
\hline\hline\\[-2ex]
& $ \pi^+$ & $\pi^-$ & $\pi^0$
&
& $ K^+$   & $K^-$   & $K^0$   & $\bar K^0$
&
& $ \eta$
\\[1ex] \hline\\[-2ex]
$\widetilde Q$-charge ${}\quad$ 
& $\ 1\ $ & $- 1$   &$\ 0\ $ 
&
& $\ 1\ $ & $ -1$   &$\ 0\ $ &$\ 0\ $ 
&
& $\ 0\ $\\[.5ex]
\hline\hline
\end{tabular}
\end{center}
\begin{center}
\begin{minipage}{.6\hsize}
{\small {\bf Table~2}: $\widetilde Q$-charges of pseudo Nambu-Goldstone bosons 
in the CFL phase.}
\end{minipage}
\end{center}

\begin{center}
\begin{tabular}{cccccccccccc}
\hline\hline\\[-2ex]
& 
\multicolumn{3}{c}{up}
&
${}\quad$
&
\multicolumn{3}{c}{down}
&
${}\quad$
&
\multicolumn{3}{c}{strange}
\\
color
& $1$ & $2$ & $3$ 
&
& $1$ & $2$ & $3$ 
&
& $1$ & $2$ & $3$
\\[1ex] \hline\\[-2ex]
$\widetilde Q$-charge ${}\quad$ 
& $\ 0\ $ & $\ 1\ $ & $\ 1\ $ 
&
& $-1$ & $\ 0\ $ & $\ 0\ $
&
& $-1$ & $\ 0\ $ & $\ 0\ $\\[.5ex]
\hline\hline
\end{tabular}
\end{center}
\begin{center}
\begin{minipage}{.6\hsize}
{\small {\bf Table~3}: $\widetilde Q$-charges of quarks  in the CFL phase.}
\end{minipage}
\end{center}

Summarizing, under the rotated electromagnetism four gluons, four
Goldstone bosons as well as four quarks are charged, as can be
read-off from the electromagnetic part of the Lagrangian (cf.~Table
1,2 and 3).  Their charges are integral multiples of the electron
charge ${\widetilde e}$. More explicitly, in the quark sector, the up
quarks of fundamental colors 2 and 3 carry charge ${\widetilde e}$,
while the down and strange quarks of fundamental color 1 carry charge
$-{\widetilde e}$. The remaining quarks are electrically neutral. The
condensates are also neutral, as expected, as the up quark of color 2
only pairs with the down quark of color 1, while the up quark of color
3 only pairs with the strange quark of color 1. Quark matter in the
CFL phase is then ${\widetilde Q}$-neutral, as in the spectrum there
are the same number of particles with positive charge ${\widetilde e}$
and negative charge $-{\widetilde e}$.  It has also been argued that
quark matter is $Q$-neutral \cite{Rajagopal:2001ff}, even in the
presence of a non-vanishing and small strange quark mass and/or
strange chemical potential. In this case, no further charge carriers
(like electrons) are needed to make the system electrically neutral.

In order to  compute the one-loop photon self  energy, it is necessary
to consider all  one-loop diagrams with two external  photon lines and
internal loops  of charged  particles.  These are  either electrically
charged quarks, pions, kaons, or gluons. In addition, we have to add a
gauge  fixing  term and  consider  the  contributions  from the  ghost
fields. The  gluons and ghosts contribution lead  to a renormalization
of the gauge coupling constant, in full analogy to the contribution to
the one-loop  photon self-energy of  the gauge sector of  the Standard
Model in vacuum as arising from  the $W_\mu ^{\pm}$ bosons and ghosts. 
Since the  gluons are heavy,  it is easy  to see that the  finite part
arising  from the  gauge boson  loops are  suppressed in  the infrared
limit.  In  contrast to  the gluons, the  charged pions and  kaons are
light.   Their  contribution  to  the photon  polarization  tensor  is
analogous to the  one in vacuum.  Apart from  a renormalization of the
coupling  constant,  they contribute  a  finite  piece  to the  photon
self-energy.  In   the  Appendix,  it  is  shown   that  their  finite
contribution is negligible in the infrared limit.  Hence, polarization
effects of the photon are  strongly dominated in the infrared limit by
pure medium  effects as arising  from the diquark condensate.   In the
following sections  we present in  full detail the computation  of the
polarization tensor due to the quarks.

\subsection{Feynman rules}\label{3-Feynman}

The computation of the one-loop photon self-energy will be done in
Euclidean space-time, using the imaginary time formalism. We first
discuss the Feynman rules for the quark propagators and vertices.
These have a complicated structure, as they mix both color and flavor
quark indices. It thus proves convenient to work in a quark basis in
which the quark propagator simplifies.

We define the color-flavor basis by the transformation~\cite{Son:2000cm}
\beq
\label{eq:2.13}
\psi_{ai} = \frac{1}{\sqrt{2}} \sum_{A=1}^9 \lambda_{ai}^A \psi^A \ ,
\eeq
where $a$ and $i$ refer to the quark flavor and color indices, respectively,
and $\lambda^A$, for $A=1, \dots,8$ are the Gell-Mann matrices, while
$\lambda^9 = \sqrt{\frac 23} 1$. 
The gap matrix can be diagonalized as
\beq
\label{eq:2.14}
\Delta^{AB} =    \delta^{AB}\,\Delta^{A}
\eeq
where $\Delta^A = k_2$ for $A=1,\ldots,8$ denotes the octet gap, 
and $\Delta^9=k_2+3 k_1$ the singlet gap.

We will use the Nambu-Gorkov formalism in the computation. The Nambu-Gorkov
fields are defined as
\begin{equation}
\label{eq:2.15}
\Psi = \left( \begin{array}{c}
               \psi \\ \psi_c
             \end{array} \right) \ , \qquad
{\bar \Psi} = \left({\bar \psi}, {\bar \psi}_c \right) \ ,
\end{equation}
where $\psi(x)$ is a Dirac spinor, while
 $\psi_c (x) = C {\bar \psi}^{T}(x)$ is the charge-conjugate spinor.

In the CFL basis the Nambu-Gorkov matrix propagator  becomes diagonal
in the $A$ index
\beq
\label{eq:2.16}
S_{AB}(K) =  \delta_{AB} \,\left(
\begin{array}{cc}
S_A^+(K) & \Xi^-_A(K) \\
\Xi^+_A(K) & S_A^-(K) 
\end{array} \right) \ ,
\eeq
where $K= (k_0, {\bf k})$ is the four momentum, and 
$k_0 = -i \omega_n = - i (2n+1) \pi T$ is a fermionic Matsubara
frequency. For massless quarks, 
\begin{mathletters}
\label{eq:2.17}
\bea
S_A^{\pm} (K) & = &\frac{\Lambda_{\bf k}^{\pm} \gamma^0
 \left(k_0 \mp \mu \pm  k \right)}
{k_0^2 - (E^A_k)^2}  +
\frac{\Lambda_{\bf k}^{\mp} \gamma^0 \left(k_0 \mp \mu \mp  k \right)}
{k_0^2 - ({\bar E}^A_k)^2}  \ , \\
\Xi_A^+ (K) & = & \gamma_5 \left\{ \frac{\Lambda_{\bf k}^- \Delta^A}
{k_0^2 - (E^A_k)^2}  + \frac{\Lambda_{\bf k}^+ {\bar \Delta}^A}
{k_0^2 - ({\bar E}^A_k)^2}  
\right\} \ , \\
\Xi_A^- (K) & = & - \gamma_5 \left\{ \frac{\Lambda_{\bf k}^+ (\Delta^A)^*}
{k_0^2 - (E^A_k)^2}  + \frac{\Lambda_{\bf k}^- ({\bar \Delta}^A)^*}
{k_0^2 - ({\bar E}^A_k)^2}  
\right\} \ , 
\eea
\end{mathletters}%
where $k = |{\bf k}|$, and
\beq
\label{eq:2.18}
\Lambda_{\bf k} ^{\pm} = 
\frac{1 \pm  \gamma_ 0 \ {\bf \gamma\cdot\hat k}}{2 } \ .
\eeq
are the positive/negative energy projectors. The energies
of particles, $E$, and antiparticles ${\bar E}$ read
\beq
\label{eq:2.19}
E^A_k  =  \sqrt{\xi_k^2 + (\Delta^{A})^2} \ , \qquad
{\bar E}^A_k  =  \sqrt{ {\bar \xi}_k^2 + ({\bar \Delta}^{A})^2} \ ,
\eeq
where in the above dispersion relations 
$\xi_k = \mu -  k$ ,
${\bar \xi_k} = \mu + k$ ,  
and $\Delta^A$ and ${\bar \Delta}^A$ are the gap and the antigap,
respectively.

From Eq.~(\ref{eq:2.8}) we can obtain the  quark-photon interaction in
the CFL basis
\beq
\label{eq:2.20}
\frac{\widetilde e}{2} \sum_{B,C=1}^9  \left[
{\rm Tr}\left( \lambda^B Q \lambda^C \right) -
{\rm Tr}\left( \lambda^C Q \lambda^B \right) \right]
{\bar \psi}^B \gamma^\mu {\widetilde A}_\mu \, \psi^C \ .
\eeq
For the complex conjugate fields the quark-photon interaction
term becomes
\beq
\label{eq:2.21}
- \frac{\widetilde e}{2} \sum_{B,C=1}^9  \left[
{\rm Tr}\left( \lambda^C Q \lambda^B \right) -
{\rm Tr}\left( \lambda^B Q \lambda^C \right) \right]
{\bar \psi}_c^B \gamma^\mu {\widetilde A}_\mu \, \psi_c^C \ .
\eeq
These terms allow us to define the Feynman rules for the
quark-photon vertices in the CFL basis.

\subsection{Traces in color-flavor space}\label{3-traces}

To one-loop order, the two diagrams given in Fig.~1
have to be evaluated. The full line in Fig.~1a (Fig.~1b) denotes 
the $S_A^{\pm}$-part ($\Xi_A^{\pm}$-part) of the quark propagator in the 
Nambu-Gorkov representation, cf.~\eq{eq:2.16}.\\

\begin{figure}
\begin{center}
\unitlength0.001\hsize
\begin{picture}(950,200)
\psfig{file=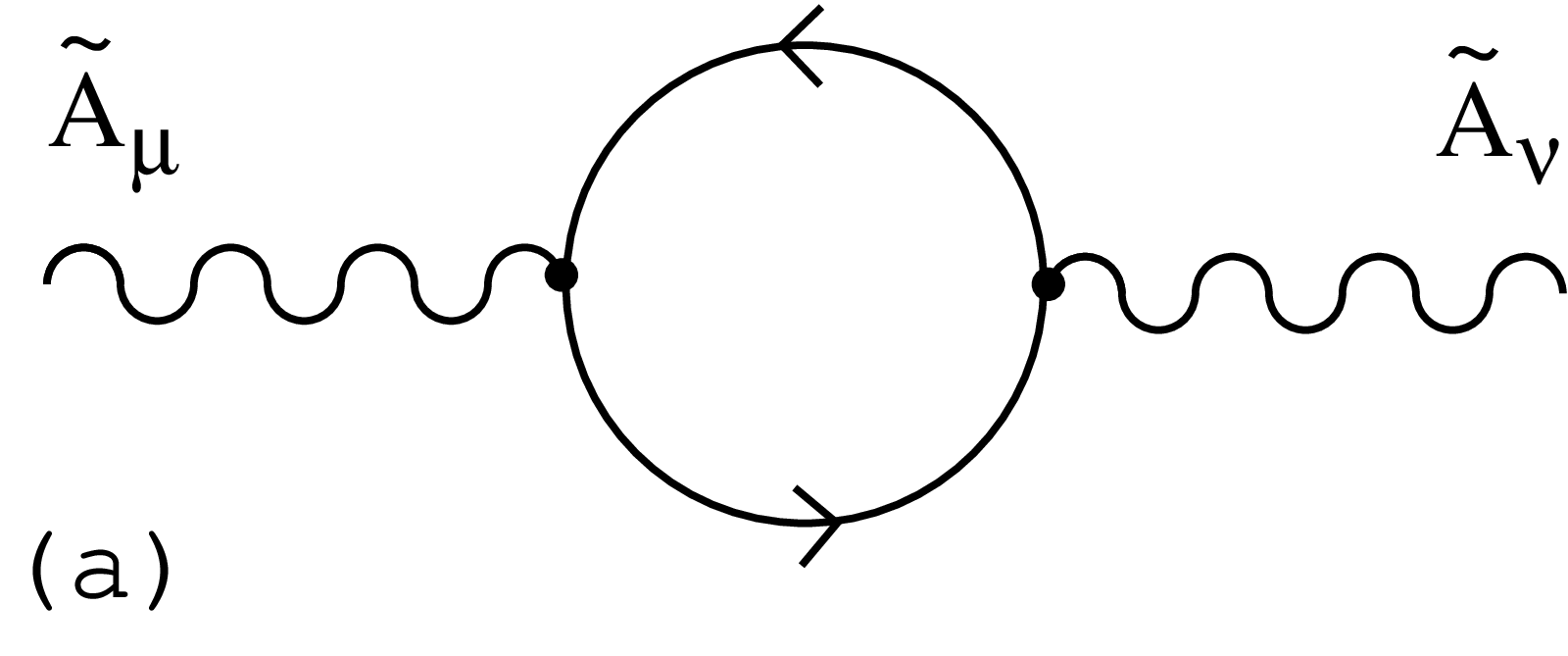,width=.45\hsize}
\hskip.05\hsize
\psfig{file=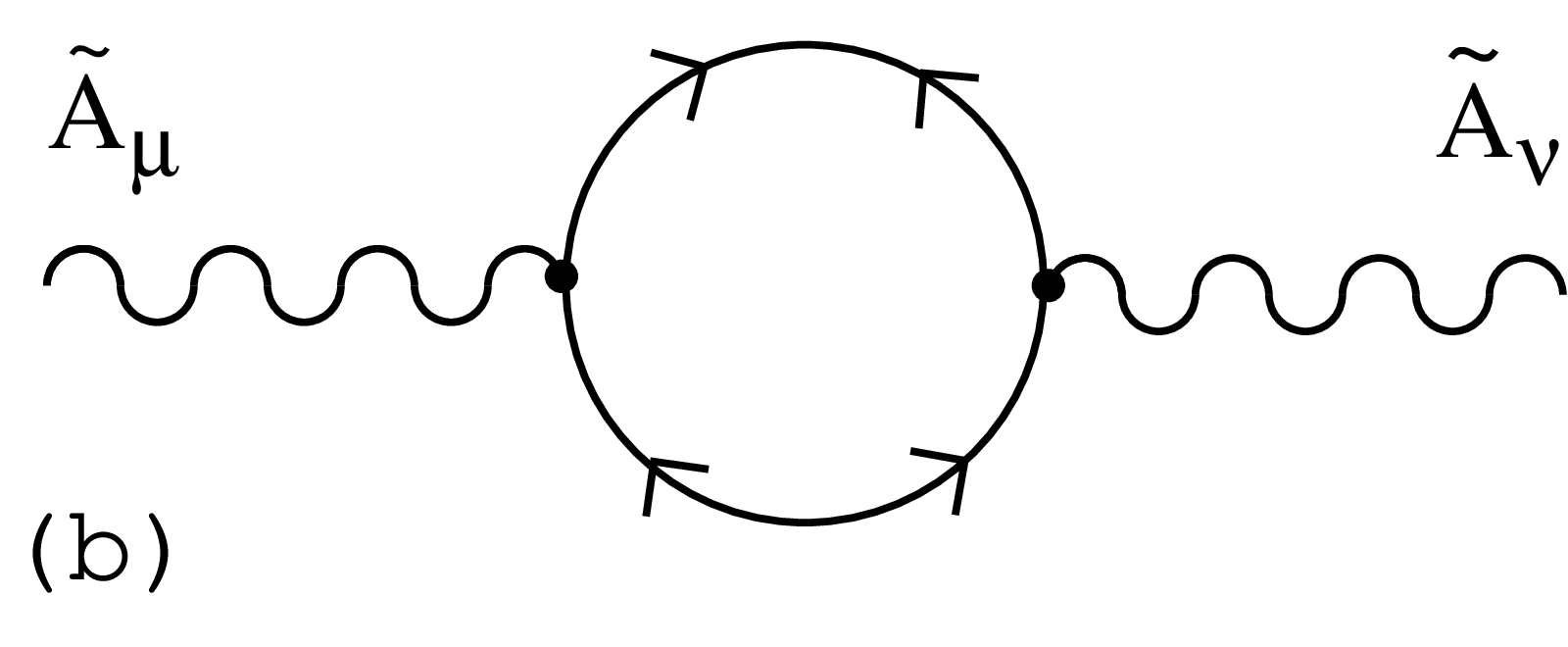,width=.45\hsize}
\end{picture}
\begin{minipage}{.95\hsize}{
    \small {\bf Figure~1}: Quark contributions to the 1-loop photon
    self energy in the superconducting phase.}
\end{minipage}
\end{center}
\end{figure}
\noindent
Using the Feynman rules of the previous subsection  one gets
\bea
\label{eq:2.22}
\Pi^{\mu \nu} (P) &  = &   \frac{\widetilde e^2}{8} 
\sum_{B,C=1}^9 \left[{\rm Tr}\left( \lambda^B Q \lambda^C \right) -
{\rm Tr}\left( \lambda^C Q \lambda^B \right) \right] 
\left[{\rm Tr}\left( \lambda^C Q \lambda^B \right) -
{\rm Tr}\left( \lambda^B Q \lambda^C \right) \right] I_{BC}^{\mu \nu} (P) \\ 
& - &    \frac{\widetilde e^2}{8} 
\sum_{B,C=1}^9 \left[{\rm Tr}\left( \lambda^B Q \lambda^C \right) 
- {\rm Tr}\left( \lambda^C Q \lambda^B \right) \right] \left[
{\rm Tr}\left( \lambda^B Q \lambda^C \right)
- {\rm Tr}\left( \lambda^C Q \lambda^B \right) \right] 
 R_{BC}^{\mu \nu} (P) \ ,
\nonumber
\eea
where 
\begin{mathletters}
\label{eq:2.23}
\bea
I_{BC}^{\mu \nu} (P) & = &  \sum_{e = \pm} \int \frac{d^4 K}{(2 \pi)^4}
{\rm Tr} \left( \gamma^\mu \, S^e_C(K) \gamma^\nu\, S_B^e(K-P) \right) \ , 
\\
R_{BC}^{\mu \nu} (P) &  = &  \sum_{e = \pm} \int \frac{d^4 K}{(2 \pi)^4}
{\rm Tr} \left( \gamma^\mu\, \Xi^e_C(K) \gamma^\nu \,\Xi_B^{-e}(K-P) \right) \
 .
\eea
\end{mathletters}%
Notice that above  we  used the standard notation of the imaginary
time formalism, so that for $T \neq 0$
\beq
 \int \frac{d^4 K}{(2 \pi)^4} \equiv T \sum_{n=-\infty}^{n = \infty} 
 \int 
\frac{d^3 k}{(2 \pi)^3} \ ,
\eeq
where the sum is over the Matsubara frequencies.

Now the $U(3)$ color-flavor traces in Eq.~(\ref{eq:2.22})
 can be easily performed, just by
realizing that the quark propagators $S_A^\pm$ and $\Xi_A ^\pm$ are
all the same for the octet, $A=1, \ldots, 8$,  while they differ
for the singlet $A=9$. In particular, the relevant traces are
\bea
\label{eq:2.24}
\sum_{B,C=1}^9 
{\rm Tr}\left( \lambda^B Q \lambda^C \right) 
{\rm Tr}\left( \lambda^C Q \lambda^B \right) M_{BC} 
& =& 
\frac{8}{9} \left(\ \ 7 M_{11} + M_{19} + M_{91} \right) 
\ , \\
\label{eq:2.25}
\sum_{B,C=1}^9 
{\rm Tr}\left( \lambda^B Q \lambda^C \right) 
{\rm Tr}\left( \lambda^B Q \lambda^C \right) M_{BC} 
& = & 
\frac{8}{9} \left(-2 M_{11} + M_{19} + M_{91} \right) \ , 
\eea
where $M_{BC}$ can be either $I_{BC}$ or $R_{BC}$. In the evaluation, 
we made use of 
$Q= \s012 \left(\lambda^3 + \frac{\lambda^8}{\sqrt{3}} \right)$. 
Finally, this leads to
\beq
\label{eq:2.28}
\Pi^{\mu \nu} (P) 
= 2\, {\widetilde e}^2\, 
\left[  I^{\mu \nu}_{11} (P) 
      + R^{\mu \nu}_{11} (P) \right] \ . 
\eeq
It is curious to note that the singlet quark propagators do not 
participate in the photon self-energy.
This is quite different to what happens in the gluon
self-energy in the CFL phase, where the one-loop terms contain
contributions from both octet quark propagators, and also from
a mixing between singlet and octet quark propagators 
\cite{Son:2000cm,Rischke:2000ra,Manuel:2001xt}.

\subsection{Sum over Matsubara frequencies}\label{3-sum}

After the  explicit evaluation of the  spinor traces and  the sum over
Matsubara frequencies  in Eqs.~(\ref{eq:2.23}), we  find an expression
for  the  one-loop  photon   polarization  tensor.  Since  the  photon
polarization tensor Eq.~(\ref{eq:2.28}) depends  only on the octet gap
$\Delta^1$,  but not  on the  singlet  gap $\Delta^9$,  we denote  the
former as $\Delta\equiv\Delta^1$  from now on. We also  drop the octet
index for  the energies,  $E \equiv E^1$,  and ${\bar E}  \equiv {\bar
  E}^1$. More explicitly, we find
\begin{mathletters}
\label{eq:2.29}
\bea
\Pi^{00} (P) & = & 
- \frac{{c\,}{\widetilde e}^2}{2}  \int \frac{d^3 \bf k}{(2 \pi)^3}
\sum_{e_1, e_2 = \pm} 
\left(1 + e_1 e_2 {\hat {\bf k}}_1 \cdot {\hat {\bf k}}_2 \right) \\
& \times & \left[ \left(\frac{1}{p_0 + E_{1} + E_{2}}
- \frac{1}{p_0 - E_1 - E_2} \right) 
\left(1 - N_1 -N_2 \right) \frac{E_1 E_2 - \xi_1 \xi_2 
- \Delta_1 \Delta_2}{ 2 E_1 E_2}  \right. \nonumber  \\
& + & \left. \left( \frac{1}{p_0 - E_{1} + E_{2}}
+ \frac{1}{p_0 + E_{1} - E_{2}} \right)
\left( N_1 - N_2 \right) \frac{E_1 E_2 + \xi_1 \xi_2 
+ \Delta_1 \Delta_2}{ 2 E_1 E_2} \right] \ , \nonumber
\\ 
\Pi^{0i} (P) & = & - \frac{{c\,}{\widetilde e}^2}{2}  
\int \frac{d^3  k}{(2 \pi)^3}
\sum_{e_1, e_2 = \pm} 
\left( e_1  {\hat  k}^i_1 + e_2 {\hat  k}^i_2 \right) \\
& \times & \left[ \left(\frac{1}{p_0 + E_{1} + E_{2}}
- \frac{1}{p_0 - E_1 - E_2} \right) 
\left(1 - N_1 -N_2 \right) \left( \frac{ \xi_2}{2E_2} -
\frac{ \xi_1}{2E_1} \right)  \right. \nonumber  \\
& + & \left. \left( \frac{1}{p_0 - E_{1} + E_{2}}
+ \frac{1}{p_0 + E_{1} - E_{2}} \right)
\left( N_1 - N_2 \right) \left( \frac{ \xi_2}{2E_2} +
\frac{ \xi_1}{2E_1} \right) \right] \ , 
\nonumber
\\ 
\Pi^{ij} (P) & = & - \frac{{c\,}{\widetilde e}^2}{2}  \int 
\frac{d^3  k}{(2 \pi)^3}
\sum_{e_1, e_2 = \pm} \left[\delta^{ij}
\left(1 - e_1 e_2 {\hat {\bf k}}_1 \cdot {\hat {\bf k}}_2 \right)
+ e_1 e_2 \left({\hat  k}^i_1 {\hat  k}^j_2 + {\hat  k}^j_1  {\hat  k}^i_2
\right) \right]
 \\
& \times & \left[ \left(\frac{1}{p_0 + E_{1} + E_{2}}
- \frac{1}{p_0 - E_1 - E_2} \right) 
\left(1 - N_1 -N_2 \right) \frac{E_1 E_2 - \xi_1 \xi_2 
+ \Delta_1 \Delta_2}{ 2 E_1 E_2}  \right. \nonumber  \\
& + & \left. \left( \frac{1}{p_0 - E_{1} + E_{2}}
+ \frac{1}{p_0 + E_{1} - E_{2}} \right)
\left( N_1 - N_2 \right) \frac{E_1 E_2 + \xi_1 \xi_2 
- \Delta_1 \Delta_2}{ 2 E_1 E_2} \right] \ ,
\nonumber  
\eea
\end{mathletters}%
where $N_i = 1/(\exp{(E_i/T)}+1)$ is the fermionic distribution
function.  The coefficient
\beq\label{Cg}
{c\,}=
\sum_{\hbox{\tiny gapped}}
\left(\0{\widetilde q_i}{\widetilde e}\right)^2 
\eeq
counts the square of the charges $\widetilde q_i$ of all gapped
quarks, in units of $\widetilde e$. In the CFL phase, ${c\,} = 4$.
The condensed notation of Eqs.~(\ref{eq:2.29}) is borrowed from
Ref.~\cite{Rischke:2000qz} and goes as follows. The indices $e_i$ take
into account the contribution of particles ($e_i = +$) and
antiparticles ($e_i = -$). All other quantities which carry a
subscript ``1'' or ``2'' have to be evaluated at the corresponding
value of the index $e_1$ or $e_2$, and for the corresponding momenta
${\bf k}_1\equiv {\bf k}$ or ${\bf k}_2 \equiv {\bf k- p}$. For
example, if $e_2 =-$, then $E_2 \equiv \sqrt{{\bar \xi}_{k-p}^2 +
  {\bar \Delta}^2} $.

It is worth emphasizing that the structure of the polarization tensor
found in Eqs.~(\ref{eq:2.29}) is very similar to the structure of the
polarization tensor of the unbroken $SU(2)_c$ gauge fields in the 2SC
phase, as computed in Ref.~\cite{Rischke:2000qz}. It suffices to
change the coupling constant and the numerical prefactor ${c\,}
\widetilde e^2 \to \frac1{2}{N_f}g^2$ to obtain the expressions given
in Eqs.~(99) of Ref.~\cite{Rischke:2000qz}, where $N_f$ is the number
of quark flavors, and the factor $\frac 12$ arises from the trace of
the $SU(3)$ generators in the fundamental representation.

\subsection{Photon self-energy at low momentum}\label{3-self}

After analytical continuation to Minkowski space-time, one can study
the behavior of the self-energy for different values of the external
frequency and momentum. Since the behavior of the polarization tensor
for the unbroken $SU(2)_c$ subgroup of the 2SC phase has been studied
in the literature, we will not present a detailed analysis here.  We
refer to the literature
\cite{Rischke:2000qz,Rischke:2000cn,Rischke:2001py} for more explicit
details.

Let us first define the longitudinal and transversal parts of the
photon polarization tensor in the usual manner,
\begin{mathletters}\label{Def-LT}
\bea
\Pi^{00} (p_0, {\bf p}) & = &
\Pi_L (p_0, {\bf p})\ , \\
\Pi^{0i} (p_0, {\bf p}) & = & 
\frac{p_0}{p}\, \frac{p^i}{p}\, \Pi_L (p_0, {\bf p}) \ , \\
\Pi^{ij} (p_0, {\bf p}) & = &  
 \left( \delta^{ij} -\frac{p^i p^j}{p^2} \right) \Pi_T (p_0, {\bf p}) 
+\frac{p^i p^j}{p^2} \, \frac{p^2_0}{p^2}\  \Pi_L (p_0, {\bf p}) \ ,
\eea
\end{mathletters}%
Below, we will only discuss the zero temperature case.  As expected,
there is no Debye or Meissner screening for the photon at vanishing
temperature, because
\begin{mathletters}\label{eq:2.31}
\bea
\lim_{p \rightarrow 0}
\Pi^{00} (p_0=0, {\bf p}) &=& 0 \ , \\
\Pi^{0i} (p_0=0, {\bf p}) &=& \Pi^{ij} (p_0=0, {\bf p}) = 0 \ .
\eea
\end{mathletters}%

In the infrared limit the photon self-energy is dominated by the quark
contribution (cf.~the Appendix).  For $p_0, p \ll \Delta$, it is
possible to compute the value of the polarization tensor
(\ref{eq:2.29}). Expanding the polarization tensor to quadratic order
in $p_0$ and $p$, and taking the limit of very large density, one
finds
\begin{mathletters}\label{eq:2.32}
\bea
\Pi_L (p_0, {\bf p})   &=&   - \widetilde\kappa\ p^2  \ , \\     
\Pi_T (p_0, {\bf p})   &=&   - \widetilde\kappa\ p^2_0\ ,  
\eea
\end{mathletters}%
where 
\beq\label{kappa}
\widetilde\kappa 
= \frac{ {c\,}}{18 \pi^2}
  \frac{\widetilde e^2\mu^2}{\Delta^2} \ .
\eeq
The coefficient ${c\,}$ has been defined in Eq.~(\ref{Cg}), $c=4$ in
the CFL phase.

The presence of the quark condensate modifies the photon dispersion
relations at low momenta. The dielectric constant of the CFL medium
becomes
\beq\label{3-epsilon}
\widetilde\epsilon 
= 1 + \widetilde\kappa 
= 1 + \frac{{2\,}}{9 \pi^2}\frac{\widetilde e^2\mu^2}{\Delta^2} \ ,
\eeq
while the magnetic susceptibility $\widetilde\lambda=1$ remains
unchanged to leading order. This is due to the fact that the CFL
condensates have zero spin and angular momentum, and hence a vanishing
magnetic moment.  The velocity $v$ of the $\widetilde Q$-photon is
given through $v^{2} =1/ {\widetilde\epsilon \widetilde\lambda}$.  In
the limit of asymptotically high densities, the gap is exponentially
suppressed $\Delta \sim \mu g^{-5} \exp{( - 3\pi^2/\sqrt{2} g)}$
\cite{Son:1999}, and therefore $\widetilde\kappa\gg 1$ and
$\widetilde\epsilon \gg 1$. As a consequence, the photon velocity is
very much suppressed when compared to the vacuum theory, $v\ll 1$.
Furthermore, the static potential created by a test charge particle,
$V(r) = {\widetilde e}/(4 \pi^2 \widetilde\epsilon\, r)$, is highly
reduced with respect to the Coulomb potential in vacuum.

For external momenta in the range $\Delta\ll p_0, p \ll \mu$, the
photon polarization tensor (\ref{eq:2.29}) reduces to the hard dense loop limit
(HDL) \cite{HTL,Manuel:1996td}, up to corrections of order ${\widetilde e}^2
\mu^2 \, \Delta/p$ \cite{Rischke:2000qz},
\beq
\label{HDL-CFL}
\Pi^{\mu \nu} (p_0, {\bf p}) = 
{\widetilde M}^2 
\left( 
- g^{\mu 0} g^{\nu 0} 
+ p_0 \int \frac{d \Omega_{\bf v}}{4 \pi} 
           \frac{v^\mu v^\nu}{p_0 - {\bf p} \cdot {\bf v} + i0^+} 
\right)  \ ,
\eeq
where
\beq\label{HDL-CFL-Debye}
{\widetilde  M}^2 = 4 \frac{{\widetilde e}^2\mu^2}{\pi^2}
\eeq 
is the corresponding Debye mass.  The
reason for why the polarization tensor reduces to the HDL result is
the following \cite{Rischke:2000qz}. For sufficiently large photon
momenta, the photon wave length is sufficiently short to resolve the
individual quarks within the Cooper pairs: the effects due to pairing
are not visible for these modes, to leading order. Consequently, one 
recovers the known leading-order result of the normal phase.

The intermediate region behavior of the functions (\ref{eq:2.29})
 with $\Delta<p_0, p \ll \mu$ has recently been
studied in \cite{Rischke:2001py}. In this region it is possible
to find analytical expressions for the imaginary part of the
self-energy, which is only non-vanishing for $p_0 \geq 2 \Delta$,
while the real part can only be studied numerically. As a result, the
photon self-energy deviates significantly from the HDL only for
frequencies $p_0 \sim \Delta$ \cite{Rischke:2001py}.

If the photon momenta is not too large $p \ll \sqrt{g} \mu$, it is
possible to show that the quark contribution dominates over that due
to charged pions and kaons in this momentum region (cf.~the Appendix).
Hence, the photon self-energy is well approximated by (\ref{HDL-CFL}).

We leave for a future project the study of electromagnetic properties
of the color superconductors at non-zero temperature $T$. However,
notice that the thermal contribution to polarization tensor
(\ref{eq:2.29}) for $p_0, p \ll \Delta$ reduces to a ``hard
superconducting loop'' \cite{Litim:2001je} and additional 
thermal corrections to the value of $\widetilde\kappa$.

\subsection{Low energy effective theory for the CFL superconductor}
\label{3-low}

The low energy effective theory for a CFL superconductor, in the absence 
of electromagnetic interactions, has been discussed in detail in
Refs.~\cite{Casalbuoni:1999wu,Son:2000cm,Rho:2000xf,Hong:2000ei,Manuel:2000wm,Zarembo:2000pj,Beane:2000ms,Miransky:2001bd,Manuel:2001xt}.
The physics is dominated by the light degrees of freedom, the
Nambu-Goldstone bosons, resulting from the spontaneous breaking of
chiral and baryon number symmetry. The low energy physics is modified
once electromagnetism is included, because the photon field is also
light.  For $p_0, p \ll 2 \Delta$, the low energy effective theory is
obtained after integrating out the heavy modes.  Including
electromagnetism, and in the chiral limit $m_q=0$, the effective
Lagrangian reads
\bea
{\cal L}     &=&   
\frac{\widetilde{\epsilon}}{2}\,
{\bf {\widetilde E}}\cdot{\bf {\widetilde E}} 
- \frac 12\, 
{\bf {\widetilde B}}\cdot{\bf {\widetilde B}} 
\nonumber \\ & & \label{eq:2.35}
+ \frac{f_\pi^2}{4} 
\left[ {\rm Tr} \left({\widetilde D}_0 \Sigma 
                      {\widetilde D}_0 \Sigma^\dagger \right) 
 - v^2_\pi {\rm Tr} \left( {\widetilde D}_i \Sigma 
                           {\widetilde D}_i \Sigma^\dagger \right) 
\right]  
+ {\widetilde e}^2\, C\, {\rm Tr} 
  \left( Q \Sigma Q \Sigma^\dagger \right)
 \ ,
\eea
where ${\bf {\widetilde E}}$ and ${\bf {\widetilde B}}$ are the in-medium
electric and
magnetic fields, respectively.
The unitary matrix $\Sigma$ contains the Goldstone fields, and
the covariant derivative acting on $\Sigma$ is
\beq
\label{eq:2.36}
{\widetilde D}_\mu \Sigma = 
\partial_\mu \Sigma 
- i {\widetilde e}\, Q\, {\widetilde A}_\mu\, \Sigma 
+ i {\widetilde e}\, \Sigma\, Q\, {\widetilde A}_\mu \ .
\eeq
In Eq.~(\ref{eq:2.35}) we have omitted the Goldstone boson associated
to the breaking of $U(1)_B$, as it does not couple to the photon.  The
values of the pion decay constant $f_\pi$ and the pion velocity
$v_\pi$ at high baryonic density have been computed from the
microscopic theory, finding
\cite{Son:2000cm,Rho:2000xf,Hong:2000ei,Manuel:2000wm,Zarembo:2000pj,Beane:2000ms,Miransky:2001bd,Manuel:2001xt}
\beq
\label{eq:2.37}
f^2_\pi = \frac{21 - 8 \ln{2}}{18} \frac{\mu^2}{2 \pi^2} \ ,
\qquad v_\pi = \frac{1}{\sqrt{3}} \ .
\eeq
In this paper we have computed the value of the dielectric constant
$\tilde\epsilon$. The last term in Eq.~(\ref{eq:2.35}) represents, to
leading order, a mass term for the charged pions and kaons. This mass
term is generated because the electromagnetic interactions represent
an explicit breaking of chiral symmetry.  The constant $C$ obeys a sum
rule, but it has not yet been computed.

At higher order in an energy expansion one could also add to
Eq.~(\ref{eq:2.35}) the Wess-Zumino-Witten term, describing anomalous
processes such as $\pi^0 \rightarrow {\widetilde \gamma} {\widetilde
  \gamma}$ \cite{Casalbuoni:2000jn,Hong:1999dk,Nowak:2001wa}.

\subsection{Discussion}\label{3-Discussion}

We briefly discuss the main physical picture which has emerged from
the present analysis. Under the rotated $U(1)$ symmetry, we have seen
that the quark charges differ from the vacuum $U(1)$ electric charges.
The in-medium charges are all integer and such that the diquark
condensates are neutral, although their components are not, in
general. In the CFL phase, the in-medium photon propagation properties
are strongly affected by the charged particles and the existence of
the diquark condensates.

\begin{figure}
\begin{center}
\unitlength0.001\hsize
\begin{picture}(900,400)
\psfig{file=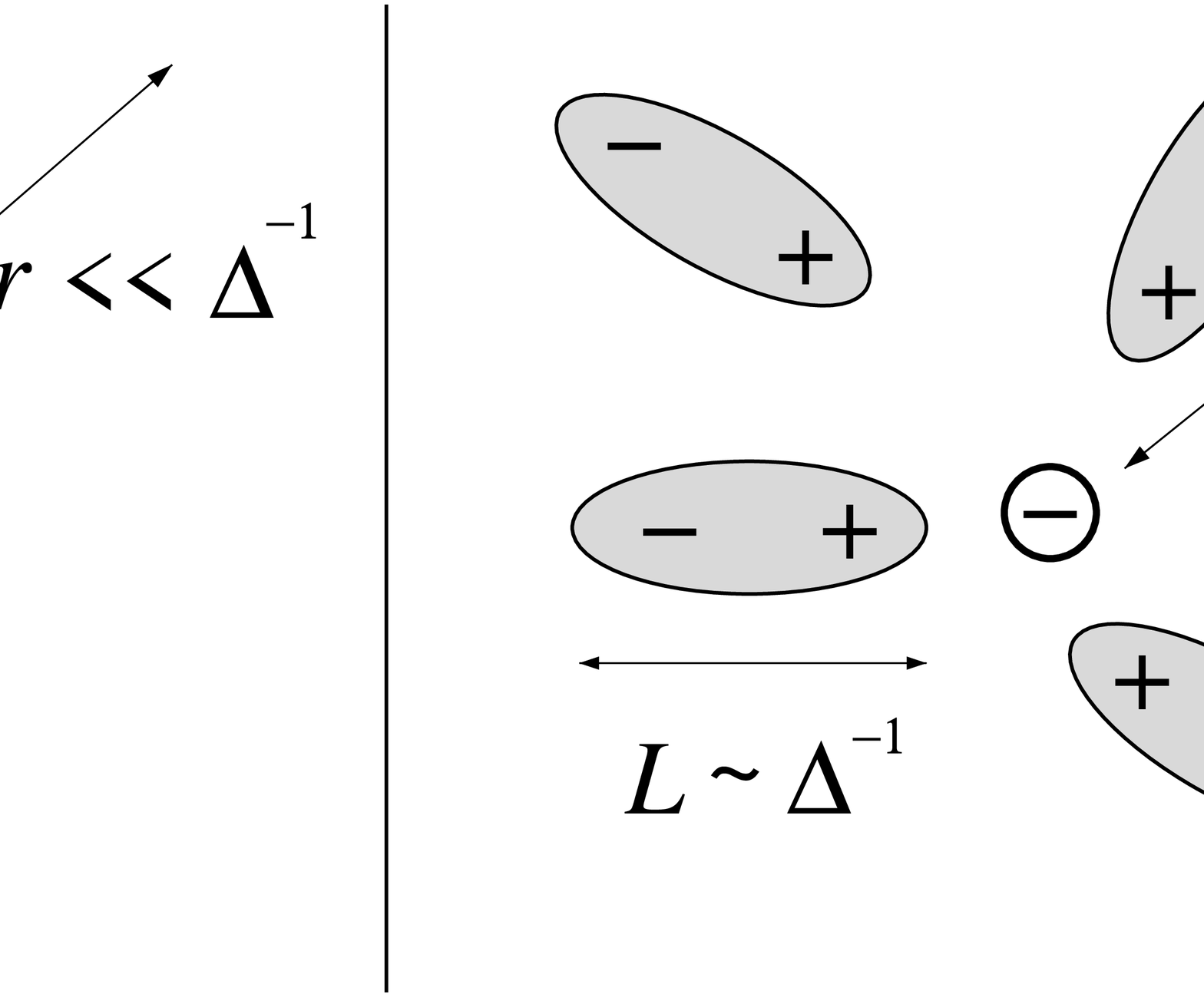,width=.9\hsize}
\end{picture}
\begin{minipage}{.9\hsize}{
\small {\bf Figure~2}: 
Screening of a negative test charge in the CFL phase. Left panel: for
length scales $r$ with $({\tilde e} \mu)^{-1}\sim r \ll \Delta^{-1}$,
the screening is dominated by effectively ungapped charged quarks.
Here, these are the up quarks of color 2 and 3. Right panel: For length
scales $r \gg \Delta^{-1}$, all quarks are bound within $\widetilde
Q$-neutral condensates. Therefore, no Debye screening takes place.
However, the condensates have an electrical dipole moment and align
themselves in an electrical field. This strongly modifies the
dielectric constant.
}
\end{minipage}
\end{center}
\end{figure}

If the wave-length associated to the photon is smaller than the mean
separation among the quarks which form the Cooper pairs, the photon
propagation is not affected by the condensate. In this regime, the
photon can resolve the components of the condensate.  Therefore the
photon self-energy matches, to leading order, the corresponding one in
the non-superconducting phase.  The electric fields acquire a Debye
mass, electric and magnetic fields are Landau damped, but there is no
static magnetic screening. The electric Debye mass $\widetilde M$
stems from all charged quarks in the CFL phase. If the photon momenta
is large enough typical quantum effects which involve the creation and
destruction of virtual pairs of charged pions and kaons would also
modify the dielectric constant associated to the electromagnetic
fields.

In turn, if the photon wave-length is large enough, larger than the
mean separation among the quarks which form Cooper pairs, it can no
longer resolve the components of the condensate. Consequently, and in
the absence of other free charge carriers (like electrons), the long
wave length photons are not Debye screened. However, the condensates
still act as electric dipoles, with total spin and angular momentum
zero. This modifies the in-medium electric properties, leading to a
strong increase of the dielectric constant $\widetilde \epsilon$.  The
in-medium magnetic permeability remains unchanged. This is due to the
fact that the condensates have vanishing magnetic moment, but would
change if the condensates had a non-vanishing spin or angular momentum
\cite{Pisarski:2000bf}. However, condensation in the lowest angular
momentum channel is favored. We conclude that the velocity of
(transverse) in-medium photons is strongly suppressed as opposed to
the vacuum.

In the static limit, the resulting picture is very simple
(cf.~Figure~2). Introducing a test charge in the superconductor would
produce a static potential.  For distances $ r \sim ({\widetilde
  e\mu})^{-1} \ll \Delta^{-1}$, the electric test charge is screened by
effectively free charge carriers; here, the up quark of color 2 and 3.
The associated static potential is given by $V(r) = {\widetilde e}
\exp (-{\widetilde M}r)/(4 \pi r)$. At larger distances, $r \gg
\Delta^{-1}$, Debye screening is no longer working because every
positively charged quark comes bundled in Cooper pairs with a
negatively charged one.  The static potential produced by a test
charge is $V(r) = \widetilde e/(4 \pi \widetilde \epsilon\, r)$.
Because $\widetilde\epsilon \gg 1$, the potential is highly reduced in
comparison to the Coulomb potential in vacuum.

Finally, we remark that all $\widetilde Q$-charged hadronic
excitations in the CFL phase, and for non-vanishing quark masses,
acquire a gap \cite{Alford:1999mk,Rajagopal:2000wf}. Therefore, the
photon cannot scatter if its energy is below the energy of the
lightest charged mode; CFL matter is thus transparent. Here, we have
computed the refraction index of CFL matter with the vacuum, given by
$\widetilde n=\widetilde \epsilon{\,}^{1/2}$. In the high density region,
the refraction index obeys $\widetilde n\gg 1$.

\section{Photon self-energy in the two-flavor color superconductor}
\label{2-flavor}

\subsection{Electromagnetic interactions}\label{2-electro}

The (spin zero) condensates of QCD with two light quark flavors
differ from those found in Eq.~(\ref{eq:2.1}). In this case one finds
\beq
\label{eq:3.1}
\langle q^{ai}_L q^{bj }_L \rangle =-
  \langle q^{ai}_R q^{bj}_R \rangle = \epsilon^{ij3} \epsilon_{ab}\,
\Delta \ ,
\eeq
where the gap itself picks up a direction in color space.
By a global color rotation this has been fixed to be the
$k=3$ direction. The original symmetries of the theory are
$SU(3)_c \otimes SU(2)_L \otimes SU(2)_R \otimes U(1)_B$, and these
are broken by the condensates to 
$SU(2)_c \otimes SU(2)_L \otimes SU(2)_R \otimes {\widetilde U}(1)_B$
\cite{Casalbuoni:2000cn}.
The group ${\widetilde U}(1)_B$ corresponds to a modified global baryon
symmetry, whose generator is ${\widetilde B} = B - \frac{2 \sqrt{3}}{3} T^8$,
where $B= \frac 13 {\rm diag}(1,1,1)$ is the standard baryon number
generator. In this section we use the standard $SU(3)$ generators, so 
that $T^8 = \frac{1}{2 \sqrt{3}} {\rm diag}(1,1,-2)$.
As a consequence of the Anderson-Higgs mechanism, five gluons acquire
masses, which are of order $\sim g \mu$, although not all of them are
equal \cite{Rischke:2000qz}.  Notice also that not all the quarks attain a gap.
In particular, the up and down quarks of fundamental color 3 are
gapless.

Exactly as it happens for the CFL superconductor, the diquark
condensates (\ref{eq:3.1}) break spontaneously the standard
electromagnetic symmetry. A local $U(1)$ symmetry, which is a
combination of the standard electromagnetic $U(1)$ and of a generator
of $SU(3)_c$, remains unbroken. In order to identify the in-medium
photon, it is convenient to use the non-linear framework.  This has
been done in full detail in
Refs.~\cite{Casalbuoni:2000cn,Casalbuoni:2000jn}. The massive ${\widetilde
  G}^8$ and massless ${\widetilde A}$ fields are \cite{Gorbar:2000ms,Casalbuoni:2000jn}
\begin{mathletters}
\label{eq:3.2}
\bea
{\widetilde G}^8_\mu 
& = & 
\ \ \, \cos{\theta_{\rm 2SC}}\, G^8_\mu 
  + \sin{\theta_{\rm 2SC}}\, A_\mu \ ,
\\
{\widetilde A}_\mu 
& = & 
- \sin{\theta_{\rm 2SC}}\, G^8_\mu 
+ \cos{\theta_{\rm 2SC}}\, A_\mu \ ,
\eea
\end{mathletters}%
where
\beq
\label{eq:3.3}
\cos{\theta_{\rm 2SC}} =  \frac{\sqrt{3} g}{\sqrt{3 g^2 +  e^2}} \ ,
\qquad  
\sin{\theta_{\rm 2SC}} = \frac{ e}{\sqrt{3 g^2 +  e^2}} \ .
\eeq 
The field ${\widetilde A}$ is the in medium photon.  Notice that the
mixing angle differ for the CFL and the 2SC superconductor.

The quark fields couple to the new photon as
\bea
\label{eq:3.4}
{\cal L}^{\rm e.m.}_{\rm quarks} =  
{\bar \psi}\, i {\widetilde D}_\mu \gamma^\mu \, \psi =  
{\bar \psi}\,
\left( i\partial_\mu-{\widetilde e}\ {\widetilde Q}\ {\widetilde A}_\mu \right) 
\gamma^\mu\,\psi \,.
\eea
The rotated gauge coupling reads $\widetilde e=e\cos{\theta_{\rm 2SC}}$, 
and the charge matrix ${\widetilde Q}$ is a matrix in
flavor$_{(2 \times 2)}$ $\otimes$ color$_{(3 \times 3)}$ space, 
\beq
\label{eq:3.5}
{\widetilde Q} = Q \otimes 1 - \frac{1}{\sqrt{3}} 1 \otimes T^8 \,. 
\eeq
In the two flavor case $Q ={\rm diag} (2/3, -1/3)$.

Some gluons are electrically charged. Using the 
definitions of Eq.~(\ref{eq:2.10}), we find 
\beq
\label{eq:3.6}
{\cal L}^{e.m.}_{\rm gluons} =
\frac 12 \, 
{\widetilde D}^{}_{[\mu } G^{+}_{\nu]} 
{\widetilde D}^{}_{[\mu } G_{\nu]}^{-} 
+\frac 12\,
{\widetilde D}^{}_{[\mu } H^{+}_{\nu]}
{\widetilde D}^{}_{[\mu } H_{\nu]}^{-} \,.
\eeq
Here, ${\widetilde D}_{\mu}X^{\pm}\equiv (\partial_\mu \pm i \frac{\widetilde
  e}{2}\, {\widetilde A_\mu}) X^{\pm}$, $X=G$ or $H$, 
and ${A}_{[\mu} B_{\nu]}\equiv A_\mu
B_\nu-A_\nu B_\mu$.  Hence, in the 2SC phase, and unlike the case in
the CFL phase, the gauge fields $G^\pm$ and $H^\pm$ of
Eq.~(\ref{eq:2.10}) have the $\widetilde Q$-charges $\s0{\widetilde e}{2}$.
The remaining gluons are electrically neutral (cf. Tab.~4).

\begin{center}
\begin{tabular}{ccccccccccc}
\hline\hline\\[-2ex]
& $G_\mu^1$ & $G_\mu^2$ & $G_\mu^3$ 
& ${}\quad$
& $G_\mu^{+}$      & $G_\mu^{-}$ 
& $H_\mu^{+}$      & $H_\mu^{-}$ 
& ${}\quad$
& $\widetilde G_\mu^8$ 
\\[1ex] \hline\\[-2ex]
$\widetilde Q$-charge ${}\quad$
& $\ 0\ $ & $\ 0\ $ &$\ 0\ $ 
&
& $\ \s012\ $ & $-\s012$ 
& $\ \s012\ $ & $-\s012$  
&
& $\ 0\ $ \\[.5ex]
\hline\hline
\end{tabular}
\end{center}
\begin{center}
\begin{minipage}{.6\hsize}
{\small {\bf Table~4}: $\widetilde Q$-charges of gluons
in the 2SC phase, and in units of 
$\widetilde e = e \cos \theta_{\rm 2SC}$.}
\end{minipage}
\end{center}

\begin{center}
\begin{tabular}{cccccccccccc}
\hline\hline\\[-2ex]
& 
\multicolumn{3}{c}{up}
&
${}\quad$
&
\multicolumn{3}{c}{down}
&
${}\quad$
&
\multicolumn{3}{c}{strange}
\\
color
& $1$ & $2$ & $3$ 
&
& $1$ & $2$ & $3$ 
&
& $1$ & $2$ & $3$
\\[1ex] \hline\\[-2ex]
$\widetilde Q$-charge ${}\quad$
& $\ \s012\ $ & $\ \s012\ $ & $\ 1\ $ 
&
& $-\s012$ & $-\s012 $ & $\ 0\ $
&
& $-\s012$ & $-\s012 $ & $\ 0\ $ \\[.5ex]
\hline\hline
\end{tabular}
\end{center}
\begin{center}
\begin{minipage}{.6\hsize}
{\small {\bf Table~5}: $\widetilde Q$-charges of quarks  in the 2SC phase.}
\end{minipage}
\end{center}

Under this rotated electromagnetism we then see that four gluons are
electrically charged, but their charges are half-integer multiples of
the electron charge ${\widetilde e}$ (cf.~Tab.~4).  
The up quarks of fundamental color
1 and 2 carry charge $\frac{\widetilde e}{2}$, while the down quarks of
color 1 and 2 carry charge $- \frac{\widetilde e}{2}$, so that the
condensates are electrically neutral. The gapless up quark of color 3
carries the charge ${\widetilde e}$, while the gapless down quark is 
neutral (cf.~Tab.~5).

In order to make the whole system electrically neutral one should add a
background of particles with negative charges, such as strange quarks
and/or electrons. The strange quarks of fundamental colors 1 and 2
carry charge $- \frac{\widetilde e}{2}$, while the strange quark of color
3 is neutral. This would be enough to make the system both $Q-$ and
${\widetilde Q}-$neutral.
  In this Section we work under the assumption that the
strange quark mass is $m_s \rightarrow \infty$, in which case the strange
quarks do not play a dynamical role. We will consider that there
is a finite density of electrons, with an associated chemical potential
$\mu_e$.

As in the previous section, we will compute the one-loop photon
self-energy due to the quark matter sector of the theory, as we are
basically interested in studying the infrared behavior of the photon
self-energy.

\subsection{Feynman rules}\label{2-Feynman}

We use the same conventions of the previous Section, and perform the
computation using the imaginary time formalism. We first give the
quark propagators in the 2SC phase. The Nambu-Gorkov propagator reads
\cite{Pisarski:2000bf}
\beq
\label{eq:3.7}
{\cal S} (K) =   \,\left(
\begin{array}{cc}
{\cal S}^+(K) & {\bf \Xi}^-(K) \\
{\bf \Xi}^+(K) & {\cal S}^-(K) 
\end{array} \right) \ ,
\eeq
where every term in the matrix is also a matrix in color,
 flavor and spinor spaces.
In particular,
\begin{mathletters}
\label{eq:3.8}
\bea
({\cal S}^{\pm} )^{ab}_{ij}(K) & = & \delta^{ab} \left(\delta_{ij}
-\delta_{i3} \delta_{j3}  \right) S^\pm (K) + 
\delta^{ab} \,\delta_{i3} \delta_{j3} \,
S^\pm_0 (K) \ , \\
({\bf \Xi}^\pm)^{ab}_{ij}(K) & = & \pm \epsilon_{ab} \, \epsilon_{ij3}
 \,\Xi^\pm (K) \ ,
\eea
\end{mathletters}%
where $a,b$ denote flavor indices, and $i,j$ are color indices.  For
massless quarks, $S^\pm$ and $\Xi^\pm$ agree with the propagators in
Eqs.~(\ref{eq:2.17}), after replacing the CFL gaps and antigaps by
their 2SC counterparts. The free propagators $S_0^\pm$, which describe
the gapless quarks, are obtained from $S^\pm$ by putting $\Delta =
{\bar \Delta}= 0$. The quark-photon vertex follows from
Eq.~(\ref{eq:3.4}).

\subsection{Traces in color and flavor space}\label{2-traces}

There are two diagrams to compute, shown in Fig.~1. Using the Feynman
rules of the previous subsection for vertices and propagators, one
gets
\bea
\label{eq:3.9}
\Pi^{\mu \nu} (P) &  = &   \frac{\widetilde e^2}{2} 
\sum_{e = \pm} \int \frac{d^4 K}{(2 \pi)^4}\,
{\rm Tr} \left( \gamma^\mu {\widetilde Q}\, {\cal S}^e(K) 
\gamma^\nu {\widetilde Q}\, {\cal S}^e(K-P) \right) \\
& - &    \frac{\widetilde e^2}{2}
\sum_{e = \pm} \int \frac{d^4 K}{(2 \pi)^4} \,
{\rm Tr} \left( \gamma^\mu  {\widetilde Q}\,
{\bf \Xi}^e(K) \gamma^\nu {\widetilde Q} \,{\bf \Xi}^{-e}(K-P) \right) 
\nonumber \ ,
\eea
where the trace above is in color, flavor and spinor spaces.
After the explicit evaluation of the color and flavor traces, one finds
\bea
\label{eq:3.10}
\Pi^{\mu \nu} (P) & = & \frac{{\widetilde e}^2}{2} 
\sum_{e = \pm} \int \frac{d^4 K}{(2 \pi)^4} \left[
{\rm Tr} \left( \gamma^\mu S^e(K) \gamma^\nu S^e(K-P) \right) +
{\rm Tr} \left( \gamma^\mu \Xi^e(K) \gamma^\nu \Xi^{-e}(K-P) \right) 
\right] \nonumber \\
& + & \frac{{\widetilde e}^2}{2}\sum_{e = \pm} \int \frac{d^4 K}{(2 \pi)^4}
{\rm Tr} \left( \gamma^\mu S^e_0(K) \gamma^\nu S_0^e(K-P) \right) \ .
\eea

We find two types of contributions to the photon self-energy.  The
first one is due to the quarks which form Cooper pairs.  This
contribution is totally analogous to the one found in the CFL phase,
Eq.~(\ref{eq:2.28}), with, however, two main differences. First, the
angles $\theta_{\rm CFL} \neq \theta_{\rm 2SC}$ are unequal, and
therefore ${\widetilde e}$ differs in the CFL and 2SC cases. Second, the
numerical factors in front of the integrals are different, which is
due to the different charge assignments of the four charged quarks in
the CFL and the 2SC phases (integer in the first case, half-integer in
the second, cf.~Tab.~2 and 5). Hence, in the 2SC phase ${c\,}=1$.  The
second contribution is due to the charged gapless up-quark. It is
totally analogous to the contribution to the photon self-energy of a
charged fermion in a dense medium with chemical potential $\mu$, and
whose value is well-known.

We will not give explicit results for evaluating the spinor traces and
the sum over Matsubara frequencies. These can be inferred from
Eqs.~(\ref{eq:2.29}), after the necessary changes in numerical factors
for the part of the polarization tensor due to the condensed quarks,
as mentioned above, and as well for the gapless quark, where
furthermore one has to put $\Delta={\bar \Delta} =0$.

\subsection{Photon self-energy at low momentum}\label{2-self}

We perform the analytical continuation to Minkowski space-time,
and study the photon polarization tensor for different values
of the external momentum and frequency. We restrict the study
to the zero temperature case.

Let us decompose the polarization tensor into longitudinal and
transverse parts as done in Eq.~(\ref{Def-LT}). Here, and in contrast
to the CFL phase, the transversal and longitudinal components consist
of two different pieces,
\begin{mathletters}\label{2SC+HDL}
\bea
\Pi_L (p_0, {\bf p})  & = &
\Pi^{\rm gap}_L (p_0, {\bf p}) + \Pi^{\rm HDL}_L (p_0, {\bf p}) \ , \\
\Pi_T (p_0, {\bf p})  & = &
\Pi^{\rm gap}_T (p_0, {\bf p}) + \Pi^{\rm HDL}_T (p_0, {\bf p}) \ .
\eea
\end{mathletters}%
The contributions $\Pi^{\rm gap}$ stem from the condensates and
correspond to the first sum in Eq.~(\ref{eq:3.10}). More explicitly,
$\Pi^{\rm gap}$ is given by Eq.~(\ref{eq:2.29}) with ${c\,}=1$ in the
2SC phase. In turn, the contribution $\Pi^{\rm HDL}$ stems from the
free charge carriers, the gapless quark and the electrons. The quark
contribution corresponds to the second sum in Eq.~(\ref{eq:3.10}). The
corresponding electron contribution reduces also to a hard dense loop.
The longitudinal and transverse components of the hard dense loop
polarization tensor are given by
\begin{mathletters}\label{HDL}
\bea
\label{eq:3.13}
\Pi^{\rm HDL}_{L} (p_0, {\bf p}) & = & {\widetilde m}^2  \left[ \frac{p_0}{2
p} \left(
 \,{\rm ln\,}\left|{\frac{p_0+ p}{p_0- p}}\right| 
-i \pi \, \Theta(p^2 -p_0^2) \right)
-1  \right] \ , \\
\Pi^{\rm HDL}_{T} (p_0, {\bf p}) & = &- {\widetilde m}^2  \, \frac{p_0^2}{2 
p^2} \left[ 1 + \frac12 \left( \frac{p}{p_0} -
\frac{p_0}{ p} \right) \, \left( {\rm ln\,} \left|{\frac{p_0+
p}{p_0- p}}\right| -i \pi \, \Theta( p^2 -p_0^2)
 \right) \, \right] \ .
\eea
\end{mathletters}%
The Debye mass due to the gapless up quarks and the electrons is 
\beq\label{2-Debye}
{\widetilde m}^2 = \frac{{\widetilde e}^2 \mu^2}{\pi^2} +
\frac{{\widetilde e}^2 \mu_e^2}{\pi^2} \ , 
\eeq
where we have supposed that
the mass of the electron can be neglected ($m_e \ll \mu_e$).

Let us first consider the deep infrared limit where both
$p_0,p\ll\Delta$. The part of the self-energy Eq.~(\ref{eq:3.10}) due
to the gapless up quark and the electron is given by Eq.~(\ref{HDL}).
The part due to the quarks forming Cooper pairs is similar to the CFL
case Eq.~(\ref{eq:2.32}). To quadratic order in $p_0,p$ the terms
$\Pi^{\rm gap}$ in Eq.~(\ref{2SC+HDL}) become
\begin{mathletters}
\bea
\Pi^{\rm gap}_L (p_0, p) & =& -\widetilde\kappa\ p^2   \ , \\
\Pi^{\rm gap}_T (p_0, p) & =& -\widetilde\kappa\ p_0^2 \ ,
\eea
\end{mathletters}%
Here, $\widetilde \kappa$ is given by Eq.~(\ref{kappa}), except that 
the numerical factor ${c\,}=1$ in the 2SC phase. The dielectric 
constant $\widetilde\epsilon$ in the 2SC medium is defined as 
\beq\label{2-epsilon}
\widetilde\epsilon 
= 1 + \widetilde\kappa 
= 1 + \frac{{1}}{18 \pi^2}\frac{\widetilde e^2\mu^2}{\Delta^2} \,.
\eeq
Notice that the dielectric constants in the 2SC and CFL phases,
Eq.~(\ref{2-epsilon}) and Eq.~(\ref{3-epsilon}) respectively, are
different. This comes about because, first, the mixing angles
$\theta_{\rm CFL} \neq \theta_{\rm 2SC}$ and thus the effective
charges are different, and second, because the charge assignments for
the four gapped quarks, and hence the numerical prefactors are 
different in the two phases.

While the HDL contribution is responsible for Debye screening,
\begin{mathletters}\label{eq:3.14}
\beq
\label{2-Screening}
\lim_{p \rightarrow 0} \Pi_L (0, {\bf p}) = - {\widetilde m}^2 
\eeq
the polarisation effects due to the condensate do not create a 
Meissner mass associated to the photon, 
\beq
\label{2-Meissner}
\Pi^{0i} (0, {\bf p})  = 
\Pi^{ij} (0, {\bf p})  = 0 \ .
\eeq   
\end{mathletters}%
For small photon momenta and frequency, Eqs.~(\ref{2SC+HDL}) tell 
us that there are two different types of screening phenomena: one
associated to the existence of  condensed quarks, responsible for 
strong polarisation effects, and one associated to the existence of
free charge carriers, responsible for Debye screening.
It is however straightforward to see that Debye screening is dominant in the
infrared limit. 

Considering the region where $p_0, p \gg \Delta$, but $p_0,p \ll \mu$,
the contribution to the photon polarization tensor due to the
condensed quarks reduces to a HDL, to leading order.  The full
polarization tensor reduces to Eq.~(\ref{HDL}), with the Debye
mass ${\widetilde m}^2$ replaced by 
\beq\label{2-DebyeHDL}
{\widetilde M}^2 = 
 \frac{2 {\widetilde e}^2\mu^2}{\pi^2} 
+\frac{{\widetilde e}^2\mu_e^2}{\pi^2}\,.
\eeq
In this momentum regime, all charged (gapped or ungapped) quarks 
contribute to the Debye mass.

\subsection{Discussion}\label{2-Discussion}

We briefly discuss the main physical picture which has emerged from
our analysis for the 2SC phase. Under the rotated $U(1)$ symmetry,
quarks have obtained integer and half-integer electric charges. All
half integer charged up and down quarks form electrically neutral
Cooper pairs. The  gapless up  quark has integer
charge. The condensates have an electric dipole moment, but no
magnetic dipole, because their total angular momentum and spin
vanishes. Therefore, photon propagation in a 2SC medium is affected
both by the free charged particles and the condensate.\\

\begin{figure}
\begin{center}
\unitlength0.001\hsize
\begin{picture}(900,400)
\psfig{file=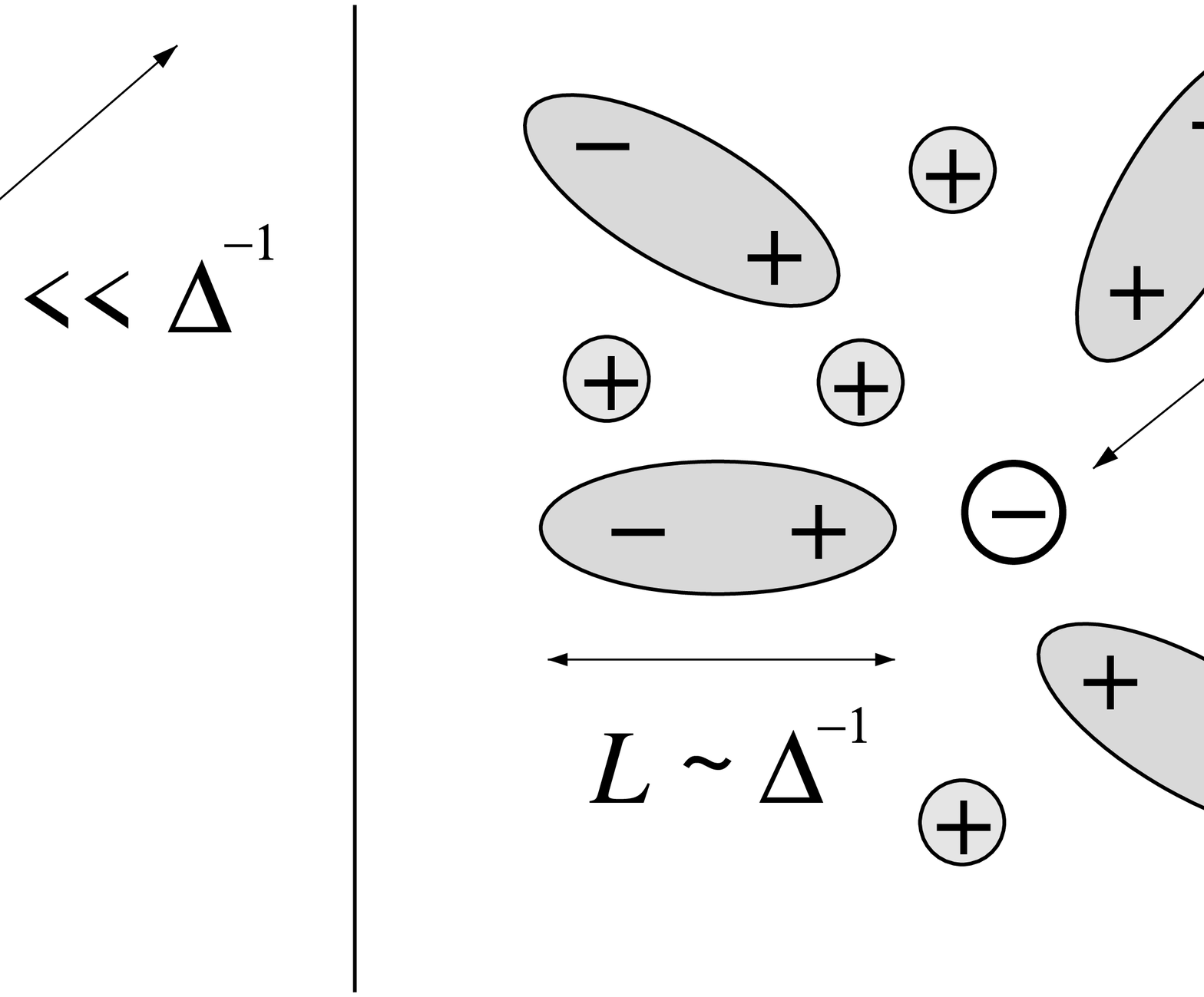,width=.9\hsize}
\end{picture}
\begin{minipage}{.9\hsize}{
\small {\bf Figure~3}: 
Screening of a negative test charge in the 2SC phase. Left panel: for
length scales $r$ with $(\widetilde e \mu)^{-1}\sim r \ll
\Delta^{-1}$, the screening is dominated by the ungapped and the
effectively ungapped charged quarks. Here, these are the up quarks of
color 1, 2 and 3, leading to the Debye mass $\widetilde M$.
Right panel: For length scales $r \gg \Delta^{-1}$, all up quarks of
color 1 or 2 are bound with down quarks of color 2 or 1 to $\widetilde
Q$-neutral condensates, while the up quark of color 3 remains gapless.
The condensates have an electrical dipole moment and modify the
dielectric constant. The gapless up quark still screen the test
charge.
}
\end{minipage}
\end{center}
\end{figure}

The photon propagation is not affected by the condensate if the photon
wavelength is smaller than the size of the Cooper pairs $\sim
\Delta^{-1}$. In this regime, the photon resolves the constituents of
the bound state and scatters with the charge carriers. Consequently,
the photon self energy equals its counterpart of the
non-superconducting phase, with photons being Landau damped, the
electric fields acquiring a Debye mass but without a static screening
of magnetic fields. The electric Debye mass ${\widetilde M}$ is of the
order $\sim \widetilde e\mu$ and stems from all charged quarks in the
2SC phase.
  
The photon can no longer resolve the diquark condensate for photon
wavelengths large compared to the size of the Cooper pairs.
Therefore, it does not scatter from the constituents. However, as in
the CFL phase, the Cooper pairs still modify the photon propagation by
inducing a large dielectric constant to the medium. The magnetic
permeability of the medium is not changed. In addition, the ungapped
charged quarks still lead to Debye screening in the long wavelength
limit. 

In the static limit, the resulting picture is as follows
(cf.~Figure~3): We introduce a negative test charge to the
superconductor which provides a static potential. At distances $r\sim
({\widetilde e \mu})^{-1} \ll \Delta^{-1}$, the electric test charge
is screened by effectively free charge carriers, in the present case
up quarks of color 1, 2 and 3. The associated static potential is
given by $V(r) = {\widetilde e} \exp (-{\widetilde M}r)/(4 \pi r)$. At
larger distances, $r \gg \Delta^{-1}$, Debye screening is less
efficient because some of the quarks are bound into Cooper pairs,
leading to a strong electric susceptibility $\widetilde \epsilon \gg
1$. The gapless quark (and the electron) still lead to Debye
screening.

\section{Summary and outlook}\label{Summary}

In this paper we have studied the propagation of electromagnetic
fields in the presence of diquark condensates at weak coupling and
high baryonic density. The condensates break spontaneously both the
non-Abelian gauge symmetry and the $U(1)$ symmetry of
electromagnetism, leaving a remaining ``rotated'' $U(1)$ symmetry
unbroken. The gauge field associated to this new symmetry, a linear
combination of the real photon and a gluon, plays the role of the
in-medium photon in the superconductor. The phenomemon is analogous to
what occurs in the electroweak sector of the Standard Model, where the
Higgs condensate breaks both the $SU(2)_L$ and $U(1)_Y$ symmetries,
but leaves the $U(1)_{\rm e.m.}$ unbroken.

We have computed the in-medium photon polarization tensor to one-loop order in
the quark fields. In both the CFL and the 2SC phase, the diquark
condensate is responsible for a large increase of the dielectric
constant of the medium, because the Cooper pairs act as strong
electrical dipoles. This effect makes the photon propagation in color
superconducting media different from the propagation in vacuum or in a
dense medium within the normal phase. 

We also found that the magnetic permeability of the medium remains as
in the vacuum theory, because the primary condensates have no magnetic
moment.  Although condensation in the $J=0$ channel is favored,
secondary condensates with non-vanishing angular momentum may form in
the 2SC or CFL phase \cite{Son:1999,Pisarski:2000bf}. In those cases,
we expect that a non-vanishing magnetic moment of the condensates
modifies the magnetic permeability of high density QCD.
 
An important qualitative difference between 2SC and CFL matter is due
to the fact that all quarks condense and acquire a gap in the CFL
phase, while in the 2SC phase, some of the charged quarks do not form
primary condensates and remain gapless. In the absence of electrons in
CFL matter \cite{Rajagopal:2001ff}, this implies for the
electromagnetic interactions that electric charges are not screened.
In turn, the remaining free charge carriers of 2SC matter always
provide a screening for electric charges.

Here, we restricted the discussion to the case of vanishing
temperature. It will be interesting to study photon propagation in
high density QCD at low temperatures, and, more generally, the related
transport properties of the medium. Most transport coefficients like
thermal or electrical conductivities, flavor diffusion or shear
viscosities are dominated by light degrees of freedom. A first step
towards a transport theory for color superconducting QCD has been made
in \cite{Litim:2001je}, where a transport equation for the gapped
quarks of 2SC matter has been discussed. We leave a more detailed
study of these questions for future investigations.
\\

\noindent{\bf Acknowledgements:}
C.~M.~thanks S.~Heinemeyer, E.~Lopez and R.~Pisarski for useful
discussions.  This work has been supported by the European Community
through the Marie-Curie fellowships HPMF-CT-1999-00404 and
HPMF-CT-1999-00391.
\vskip-.5cm

\section*{Appendix: Photon self energy and charged scalar mesons}
\setcounter{section}{1}
\renewcommand{\theequation}{\Alph{section}.\arabic{equation}}

Here, we compute the contributions to the photon polarization tensor
due to charged scalar mesons in the CFL phase.  For the discussion in
the main text, we are interested in its finite part. We show that
contributions from charged scalar mesons can be neglected, compared to
the quark contribution, in the infrared limit.  We restrict the
discussion to the case of vanishing temperature.  The main computation
is fully equivalent to the computation of the one-loop polarization
tensor for scalar QED, which we recall as well (see, for example,
Ref.~\cite{Schubert,Itzykson:1980rh}).

\subsection*{1. One-loop polarization tensor of scalar QED}

Consider electrically  charged scalar fields with  charge ${\tilde e}$
and  mass $m$.  To one-loop  order, two  diagrams with  charged scalar
fields  propagating   within  the  loops  contribute   to  the  photon
polarization  tensor. In Euclidean  space-time, it  can be  written as
\cite{Itzykson:1980rh}
\beq\label{scalarQED-Def}
\Pi^{\rm qed}_{\mu\nu}(P,m^2)=
{\tilde e}^2
\int\0{d^dK}{(2\pi)^d}
\left[
\0{-2\delta_{\mu\nu}}{K^2+m^2}
+
\0{(2P+K)_\mu\,(2P+K)_\nu}{((P+K)^2+m^2)(K^2+m^2)}
\right]\,.
\eeq
Here,  $P^\mu \equiv  (p_0,{\bf p})$.   The first  term in  the square
brackets  stems   from  the   tadpole  diagram.   Using   the  Feynman
parametrization  for  the loop  integrals,  and  performing a  partial
integration, we end up with the following expression:
\beq\label{def}
\Pi_{\mu\nu}(P,m^2)=
\left({P_\mu P_\nu}-{P^2}\delta_{\mu\nu} \right)
\widehat\Pi(P^2,m^2) \,,
\eeq
where
\beq\label{full}
\widehat\Pi(P^2,m^2)=
\0{{\tilde e}^2}{(4\pi)^{d/2}}
\Gamma\left(2-\s0{d}{2}\right)
\int_0^1 dx
(1-2x)^2 
\left(x(1-x)P^2+m^2\right)^{\s0{d}{2}-2}\,.
\eeq
The  difference   between  scalar  and  spinor  QED   amounts  to  the
replacement $(1-2x)^2\to 8x(1-x)$ in the integrand of Eq.~(\ref{full})
\cite{Itzykson:1980rh}.  Notice  that the dimensions  of $\widehat\Pi$
and  $\Pi_{\mu\nu}$ are different.   From Eq.~(\ref{full}),  and after
analytical  continuation  to  Minkowski  space,  we  deduce  that  the
polarization tensor has a cut  and acquires an imaginary part for $P^2
\ge 4 m^2$ in $d\le 4$ dimensions.  Within dimensional regularisation,
and after expanding in  $\epsilon=(4-d)/2$, we find both the divergent
and finite parts of the one-loop self-energy:
\bea
\widehat\Pi(P^2,m^2)&=&
\0{{\tilde e}^2}{(4\pi)^{2}}
\left[
\0{1}{3\epsilon}
+\0{1}{3}(\gamma-\ln 4\pi)
+\0{1}{3}\ln \0{m^2}{\Lambda^2}
\right.
\nonumber \\ \label{sqed-4d} &&
\quad\quad\quad\left.
-\089 -\083 \0{m^2}{P^2}
+\023 \left(1+4\0{m^2}{P^2}\right)^{3/2}
  {\rm arccoth}\sqrt{1+4\0{m^2}{P^2}}
+{\cal O}(\epsilon)
\right]
\eea
Here,  $\gamma$  is  the  Euler  constant, and  $\Lambda$  denotes  an
arbitrary renormalization scale.  Let  us denote the renormalized part
of Eq.~(\ref{full})  as $\widehat\Pi_R(P^2,m^2) = \widehat\Pi(P^2,m^2)
- \widehat\Pi(P=0,m^2)$.    

We  are  interested in  computing  the  finite  part of  the  one-loop
self-energy   in   the    infrared   limit.    For   $P^2/m^2\ll   1$,
Eq.~(\ref{full}) can be Taylor-expanded, and we find
\beq
\label{low-momentum}
\widehat\Pi_R(P^2,m^2)= 
\0{{\tilde e}^2}{(4\pi)^{2}}
\left[\0{1}{30}\0{P^2}{m^2}-\0{1}{420}\0{P^4}{m^4}+\ldots\right]
\eeq
In the  static limit where $P^2  = - p^2$, the  one-loop correction to
the photon self-energy modifies the dielectric constant of vacuum as
\beq
\epsilon = 1 - \frac{{\tilde \alpha}}{120 \pi} \frac{ p^2}{m^2} \ ,
\eeq
where  ${\tilde \alpha}={\tilde  e}^2/(4\pi)$. This  result  was first
noticed  by  Uehling  in  the  context  of  spinor  QED
\cite{Itzykson:1980rh}.  The  physical
interpretation of  this result is the  following: quantum fluctuations
create virtual  pairs of positively and  negatively charged particles,
which  act  as  electric  dipoles  and contribute  to  the  dielectric
constant of the  vacuum.  The quantum vacuum is  not really empty, but
full of virtual  pairs of particles which are  continuosly created and
destroyed.

From  Eq.~(\ref{full}) one  can as  well consider  the  opposite limit
$-P^2/m^2 \gg 1$.  The finite part of the  renormalized Green function
then reduces to
\beq\label{sqed-large}
\widehat\Pi_R(P^2,m^2)= 
\frac{{\tilde \alpha}}{12 \pi} 
\left[-  \ln{\frac{|P^2|}{m^2}} +\frac 83 +\ldots \right] \ .
\eeq
%

\subsection*{2. One-loop polarization tensor of mesons in the CFL phase}

Now we turn  to the computation of the  photon polarization tensor due
to the charged pions and kaons  in the CFL phase.  We denote the meson
masses as $m_{\pi^\pm}$  and $m_{K^\pm}$.  They are not  yet known, as
the contribution arising from the electromagnetic interactions has not
been computed.   Nevertheless, we  know that their  masses have  to be
$m_s  \ll 2  \Delta$  (with $s=\pi^\pm,\,K^\pm$),  as otherwise,  they
would    not    be    stable.     A   rough    estimate    given    in
Ref.~\cite{Manuel:2001xt}, and dimensional  analysis suggest that $m_s
\sim  {\tilde  e}  \Delta$  in  the  chiral  limit.   Furthermore,  at
vanishing baryonic density,  the mesons travel at the  speed of light. 
In  contrast, within  CFL  superconducting matter,  their velocity  is
$v_\pi < 1$, as can be read-off from Eq.~(\ref{eq:2.12}).

We can  infer the one-loop  contribution to the photon  self-energy in
the CFL phase from the vacuum  result of scalar QED, by rescaling both
the   spatial   derivatives   and   the  spatial   gauge   fields   as
$\partial_\mu\to    \tilde\partial_\mu    \equiv   (\partial_0,v_\pi\,
\partial_i)$ and  $A_\mu\to \tilde A_\mu  \equiv (A_0,v_\pi \,  A_i)$. 
Using the Feynman rules as implied by Eq.~(\ref{eq:2.12}), we find that for
a charged Goldstone boson field with mass $m_s$
\beq\label{scalarQED-CFL}
\widetilde\Pi_{\mu\nu}(P)= 
\frac{{\tilde e}^2}{(v_\pi)^{d-1}}
\int\0{d^d{\tilde K}}{(2\pi)^d}
\left[
\0{-2\delta_{\mu\nu}}{{\tilde K}^2+m_s^2}
+
\0{(2{\tilde P}+{\tilde K})_\mu\,(2{\tilde P}+{\tilde K})_\nu}{(({\tilde P}+
{\tilde K})^2+m_s^2)({\tilde K}^2+m_s^2)}
\right]
\eeq
with Euclidean momenta ${\tilde P}^\mu = (p_0, v_\pi\, {\bf p})$.  The
explicit  evaluation  of  Eq.~(\ref{scalarQED-CFL}) is  simplified  by
noticing      that      Eq.~(\ref{scalarQED-CFL})      reduces      to
Eq.~(\ref{scalarQED-Def}), the  polarization tensor for  scalar QED in
vacuum:
\beq
\widetilde\Pi_{\mu\nu}(p_0,{\bf p})= 
(v_\pi)^{-3}\, \Pi^{\rm qed}_{\mu\nu}(p_0,v_\pi\,{\bf p})\,.
\eeq
Finally,  we  undo the  rescaling  and equate  $A_\mu\Pi_{\mu\nu}A_\nu
\equiv \tilde  A_\mu\widetilde\Pi_{\mu\nu}\tilde A_\nu $,  in order to
find the following components of $\Pi_{\mu\nu}$:
\begin{eqnarray}\label{Pi-Meson1}
\Pi^{00}(P) & = & 
\frac{p^2}{v_\pi} \,
\widehat\Pi_R(\tilde P^2,m_s^2) 
\ , \\ \label{Pi-Meson2}
\Pi^{0i}(P)  & = & 
- \frac{p^0\,p^i}{v_\pi} \,
\widehat\Pi_R(\tilde P^2,m_s^2) 
\ , \\ \label{Pi-Meson3}
\Pi^{ij}(P)  & = & 
\left( \left(\frac{p_0^2}{v_\pi}+ v_\pi  p^2 \right) \delta^{ij}
-v_\pi\, p^i\, p^j\right)
\widehat\Pi_R({\tilde P}^2,m_s^2) 
\end{eqnarray}
This polarization tensor obeys  the Ward identity $P_\mu \Pi^{\mu \nu}
=0$.  Decomposing Eqs.~(\ref{Pi-Meson1})  --  (\ref{Pi-Meson3}) as  in
Eq.~(\ref{Def-LT}), we find
\bea\label{Pi-MesonL}
\Pi_L(P)& = & 
\frac{p^2}{v_\pi} \,
\widehat\Pi_R(\tilde P^2,m_s^2) 
\ , \\ \label{Pi-MesonT}
\Pi_T(P)& = & 
\left(\frac{p_0^2}{v_\pi}+ v_\pi  p^2 \right)
\widehat\Pi_R({\tilde P}^2,m_s^2) \,,
\eea
the longitudinal and transversal part of the mesonic contribution to
the photon self-energy.

\subsection*{3. Discussion}

The infrared  limit of this  polarization tensor can be  obtained from
Eq.~(\ref{low-momentum}).  At  this stage, it  is easy to see  why the
low momentum contribution of charged  pions and kaons in the CFL phase
is negliglible  as compared to  the quark contribution as  computed in
Sect.~\ref{3-self}. Consider  the longitudinal polarization  tensor in
the  static limit.  From  Eq.  (\ref{Pi-MesonL}),  we deduce  that the
leading  contribution  from  the  charged  mesons for  $p\ll  m_s$  is
proportional to $(\tilde e^2 p^4  v_\pi)/ m^2$. For $p\ll \Delta$, the
quarks contribute  proportional to $(\tilde e^2  p^2 \mu^2)/\Delta^2$. 
Comparing  the   two  expressions,  it  is  obvious   that  the  meson
contribution is  always negligible, because $\mu/\Delta  \gg 1$, while
$p/m_s\ll 1$.  Therefore, as pointed  out in the main text, the photon
polarization  tensor is  fully dominated  by  the quark  loops in  the
infrared limit.

Let us now discuss the intermediate regime where $\Delta \ll p_0,p \ll
\mu$.  We consider the longitudinal polarization tensor in the static
limit.  As has been shown in Eqs.~(\ref{HDL-CFL}) and
(\ref{HDL-CFL-Debye}), the quark contribution leads to a Debye mass
$\Pi^{\rm quark}_L=(4\tilde e^2\mu^2)/\pi^2$.  From
Eqs.~(\ref{sqed-large}) and (\ref{Pi-MesonL}), we deduce that the
meson contribution is given by
\beq
\Pi^{\rm meson}_L=
\frac{{\tilde e^2} p^2}{24 \pi^2 v_\pi} 
\left[-  \ln{\frac{p^2 v^2_\pi }{m^2}} +\frac 83 \right]\,.
\eeq
Here, we assumed that $ p \gg m_s/v_\pi$. While the explicit values of
the  meson  masses  are  unknown,  their  scale  is  set  by  the  gap
$m_s\propto  \Delta$.   In  this   case,  we  can  perform  a  leading
logarithmic  approximation  in the  non-Abelian  gauge coupling.   For
sufficiently small $g$, the  logarithm is dominated by the exponential
suppression  of  the  gap  \cite{Son:1999}.   To  leading  logarithmic
accuracy,  and in the  momentum regime  considered, we  have $\ln[(p^2
v^2_\pi)/(m^2)]=(6\pi^2)/(\sqrt{2} g)  +{\cal O}(\ln g)+{\cal  O}(1)$. 
Consequently,  $|\Pi^{\rm meson}_L|=  ({\tilde  e^2} p^2)/(4  \sqrt{2}
v_\pi\, g)$ to leading order in $g$.  Although the contribution itself
is  enhanced by  $1/g$, it  remains subleading  compared to  the quark
contribution   for   $p^2\ll   (16\sqrt{2}v_\pi\,  g\,\mu^2)/\pi^2$.   
Dropping   irrelevant  numerical  factors,   the  bound   reads  $p\ll
\sqrt{g}\mu$.  Notice  that the $\tilde  e$ has disappeared  from this
estimate,  because the leading  logarithmic approximation  is entirely
due  to $g\to 0$.   We emphasize  that this  estimate is  a worst-case
analysis.   For larger  $g$, the  logarithm is  less dominant  and the
boundary where meson and  quark contributions are comparable is pushed
towards higher momenta.



\end{document}